\documentclass[11pt]{article}
\usepackage[letterpaper, left=1in, top=1in, right=1in, bottom=1in,nohead,includefoot]{geometry}
\usepackage{color,charter,graphicx,natbib,here,setspace,multirow,amsmath,amssymb,amsfonts,enumitem,placeins}

\usepackage[utf8]{inputenc}

\usepackage[dvipsnames]{xcolor}
	\usepackage[pdftex]{hyperref}
	\hypersetup{colorlinks,
	citecolor=NavyBlue  
	}
	
\newcommand*{\missingreference}{\color{red}{??}}
\newcommand*{\missingcitation}{\color{red}{??}}
\makeatletter
\def\@setref#1#2#3{%
   \ifx#1\relax
    \protect\G@refundefinedtrue
    \nfss@text{\reset@font\missingreference}%
    \@latex@warning{Reference `#3' on page \thepage \space
              undefined}%
   \else
    \expandafter#2#1\null
   \fi}
\def\@citex[#1]#2{\leavevmode
   \let\@citea\@empty
   \@cite{\@for\@citeb:=#2\do
     {\@citea\def\@citea{,\penalty\@m\ }%
      \edef\@citeb{\expandafter\@firstofone\@citeb\@empty}%
      \if@filesw\immediate\write\@auxout{\string\citation{\@citeb}}\fi
      \@ifundefined{b@\@citeb}{\hbox{\reset@font\missingcitation}%
        \G@refundefinedtrue
        \@latex@warning
          {Citation `\@citeb' on page \thepage \space undefined}}%
        {\@cite@ofmt{\csname b@\@citeb\endcsname}}}}{#1}}
\makeatother

\bibpunct{(}{)}{;}{a}{}{,} 

\def\figfile{jpg}
\def\figdir{figurescolorjpg} 

\def\a{\mbox{\boldmath$a$}}

\def\q{\mbox{\boldmath$q$}}
\def\u{\mbox{\boldmath$u$}}
\def\U{\mbox{\boldmath$U$}}
\def\z{\mbox{\boldmath$z$}}

\def\h{\mbox{\boldmath$h$}}
\def\H{\mbox{\boldmath$H$}}
\def\A{\mbox{\boldmath$A$}}

\def\C{\mbox{\boldmath$C$}}
\def\U{\mbox{\boldmath$U$}}

\def\F{\mbox{\boldmath$F$}}
\def\V{\mbox{\boldmath$V$}}

\def\R{\mbox{\boldmath$R$}}
\def\I{\mbox{\boldmath$I$}}
\def\zero{\mbox{\boldmath$0$}}
\def\one{\mbox{\boldmath$1$}}
\def\x{\mbox{\boldmath$x$}}
\def\y{\mbox{\boldmath$y$}}
\def\b{\mbox{\boldmath$b$}}

\def\bmu{\mbox{\boldmath$\mu$}}
\def\bbeta{\mbox{\boldmath$\beta$}}
\def\bPhi{\mbox{\boldmath$\Phi$}}

\def\btheta{\mbox{\boldmath$\theta$}}

\def\W{\mbox{\boldmath$W$}}

\def\bomega{\mbox{\boldmath$\omega$}}

\def\m{\mbox{\boldmath$m$}}

\def\h{\mbox{\boldmath$h$}}

\def\mD{\mathcal{D}}
\def\mH{\mathcal{H}}
\def\seq#1#2{#1{:}#2}
\def\eqno#1{eqn.~(\ref{eq:#1})}

\begin{document}

\title{Dynamic Mixed Frequency Synthesis \\for Economic Nowcasting}  
\author{Kenichiro McAlinn\thanks{Booth School of Business, University of Chicago,  Chicago, IL 60637. {\scriptsize  Email: kenichiro.mcalinn@chicagobooth.edu} }
}
 
\maketitle\thispagestyle{empty}\setcounter{page}0

\begin{abstract}
  
We develop a Bayesian framework for dynamic modeling of mixed frequency data to nowcast quarterly U.S. GDP growth.
The introduced framework utilizes foundational Bayesian theory and treats data sampled at different frequencies as latent factors that are later synthesized, allowing flexible methodological specifications based on policy goals and utility.
Time-varying biases and inter-dependencies between data sampled at mixed frequency  are learnt and effectively mapped onto easily interpretable parameters, which can then be used to improve forecasts/nowcasts and decisions.
A macroeconomic study of nowcasting quarterly U.S. GDP growth using a number of monthly economic variables demonstrates improvements in terms of nowcast performance and interpretability compared to several methods standard in the literature.
The  study further shows that incorporating information during a quarter markedly improves the performance in terms of both point and density nowcasts, while providing insights into the economy.

\end{abstract}

\newpage

\section{Introduction \label{sec:intro} }

Utilizing data sampled at different frequencies to improve inference and forecast has garnered interest in many fields of application due to its practical utility and necessity.
For macroeconomics, in particular, the need for continuous updating of current economic situations to make timely, informed policy decisions has bolstered this interest, especially in the context of nowcasting \citep[see, e.g.,][for a survey]{banbura2012nowcasting}.
The problem derives from certain key indicators having limited sampling frequencies due to practical limitations.
For example, U.S. GDP is sampled only at the quarterly level, while many other economic variables (including inflation, unemployment, etc.) are sampled every month, with financial variables sampled at much higher frequency.

Conforming to the lowest frequency indicator-- by, for example, averaging higher frequency data, a standard approach-- have caused significant limitations for decision and policy makers.
Notably, recent economic crises and shocks have made clear that activities between samples, reflected in data sampled at higher frequencies, provide crucial information to intervene/adjust/revise nowcasts and policy.
Limitations in conventional modeling, and the necessity to incorporate data from mixed frequencies have promoted research and methodological development in the field of economics and statistics.

There are two main strands of development to model mixed frequency data: mixed data sampling (MIDAS) regression and mixed frequency vector auto-regressive models (MF-VAR), though these are certainly not the only methods developed and used in the literature or in practice.
Notably, bridge equations \citep{baffigi2004bridge,diron2008short}, where each high frequency data is related to the target data by averaging, and shrinkage methods \citep{carriero2015realtime}, where a potentially large set of high frequency data are made sparse using model selection via shrinkage, have been proven to be potent in nowcasting.
Here, however, we focus on the two main strands of development in the literature.
MIDAS, proposed by \cite{ghysels2004midas,GhyselsSantaClaraValkanov2005,GhyselsSantaClaraValkanov2006} in the context of financial econometrics, uses distributed lags on the high frequency data to ensure parsimonious specifications to {\it adjust} the data to the lower frequency regressand.
The practical appeal of MIDAS is its flexibility with the specification of the distributed lag.
Due to this, many different  specifications have been proposed (starting with the Beta and Almon lag of the seminal paper) for different applications.
The MIDAS approach was later extended for macroeconomic applications by \cite{ClementsGalvao2008,ClementsGalvao2009}, with the goal to nowcast U.S. GDP output growth using monthly economic variables \citep[see, e.g.,][for other applications]{marcellino2010factor,AndreouGhyselsKourtellos2010,galvao2013changes,guerin2013markov,lahiri2013nowcasting,aastveit2017density,casarin2017uncertainty}.
MF-VAR, proposed by \cite{mariano2003new,mariano2010coincident}, is a VAR model operating at the highest frequency, interpolating the low frequency data during samples, specifically developed and applied to macroeconomic studies \citep{camacho2013mixed,Aastveit2014,schorfheide2014identifying,schorfheide2015real,foroni2015markov}.
The assumption behind MF-VAR is that there is a latent process behind the data that can be captured with the high frequency process.
Interpolating the lower frequency data, thus yields a realization from this process.
The two competing methods have strengths and weaknesses, compared and discussed in \citet{kuzin2011midas} and \citet{foroni2014comparison} for macroeconomic studies.

Our approach fundamentally differs from both MIDAS and MF-VAR.
Rather than attempt to jointly model different frequencies at once, whether it be in a single regression (MIDAS) or a system of equations (MF-VAR), we propose a two stage Bayesian approach that is consistent with foundational Bayesian theory.
Bayesian mixed frequency synthesis (MFS) ``projects" data sampled at higher frequencies to the low frequency data of interest, treating the projections as providers of information in the form of prior distributions on the low frequency data of interest.
These priors are then synthesized as latent distributions-- biases and inter-dependencies among and between projections learnt and calibrated-- using the recently developed prior-posterior updating scheme of Bayesian predictive synthesis \citep[BPS:][]{McAlinn2016}.
By utilizing the BPS framework, MFS constructs a  coherent Bayesian framework for mixed frequency modeling.
MFS additionally provides the policy maker with significant flexibility; dictating how to construct the latent priors for the mixed frequency data, and how to synthesize these prior information.
In particular, the BPS framework allows for MFS to incorporate dynamic learning with time varying parameters on the mixed frequency data, something that has not been developed previously for MIDAS or MF-VAR.


Section~\ref{MFS} details the foundations of MFS and the general framework on which we will examine mixed frequency analysis.
Section~\ref{nowGDP} lays out the data and specifications used in the analysis in Section~\ref{resGDP}, where we examine the  nowcasting results using MFS and other methods in the macroeconomic study.
Additional comments in Section~\ref{sum} conclude the paper.



\section{Bayesian mixed frequency synthesis \label{MFS}}

Bayesian MFS is a two stage approach.
The first stage, which we call {\it frequency projection}, projects the individual higher frequency data to the low frequency data of interest.
The second stage, which we call {\it frequency synthesis}, then synthesizes the projected information from the first stage onto a forecast, or nowcast, of the low frequency data.
A key departure from the literature is that we treat the ``projections'' from each high frequency data as prior distributions on some latent state (of the economy, represented by that data), providing information to the decision maker in the form of a density.
By treating these projected information as priors, we can then synthesize them to produce a posterior distribution using the Bayes theorem in \cite{McAlinn2016}.
The resulting posterior distribution will include all of the information from-- multiple sources of-- the high frequency data, as a set of latent states; with the biases and latent dependencies learnt.
We will first discuss the frequency projection step, and then proceed to discuss frequency synthesis using the BPS framework.

Throughout, the low frequency data of interest is denoted as $y_t$ and the high frequency data as $z_{t+m/n}$, where $t=1{:}T$ and $m=0{:}n$ is the inter frequency datapoint between $t$ and $t+1$.
For example, if $y_t$ is sampled at the quarterly frequency and  $z_{t+m/3}$ at the monthly frequency,  $m$ will be from $0:3$ for each $t$, where $m=0$ is on the same frequency as the quarterly data, $m=1$ is the first month in that quarter, and so on.

\subsection{Frequency projection}

The first step requires the decision maker to specify how she believes the high frequency data should be projected on to the low frequency data of interest.
An illustrating example, though certainly not limited to, is a simple linear regression,
\begin{equation}\label{eq:proj1}
	y_{1:t}=\bbeta'\z_{1:t}+\epsilon, \quad \epsilon_t\sim N(0,\nu)
\end{equation}
where $\bbeta$ is a $p\times 1$ vector of regression coefficients, and $\z_{1:t}$ is a $p\times t$ matrix where each column is a $p\times 1$ vector of different lagged values of $z_{t+m/n}$. 
For example, if $z_{t-1+m/3}$ with $m=1:3$ for $t$, then the monthly data within the past quarter is used as regressors.
As a result of eqn.~(\ref{eq:proj1}), the decision maker obtains, for each $t=1{:}T$, a predictive distribution for $t+1$; $N(x_{t+1}|\bbeta'\z_{t+1},\nu)$, where $x_{t+1}$ is the predictive value of $y_{t+1}$ using the high frequency data vector, $\z_{t+1}$.
The resulting predictive distributions are ``projected" information of $\z_t$ onto $y_t$, and thus a projection of the high frequency data onto the low frequency data.
This, in a way, is a filtering process of taking a larger dimension data and extracting the relevant information to the quantity of interest.
A relevant analogy here is to say that the projected information formulates a latent state on $y_{1:t}$, a reduced form representation of the underlying data generating process.
We note that, if the projection model is correctly specified, i.e., the data generating process is a linear combination of $\z_t$  on $y_t$, all of the relevant information from $\z_t$ concerning $y_t$ is contained in $N(x_{t}|\bbeta'\z_t,\nu)$, giving the decision maker a perfect prediction of $y_t$.
However, if the projection methods are not correctly specified, $N(x_{t}|\bbeta'\z_t,\nu)$ will be biased.
In such a complicated system as the economy, we cannot expect one, or even a few, higher frequency data to perfectly capture $y_t$.
Thus, in all relevant applications, $N(x_{t}|\bbeta'\z_t,\nu)$ will not only be biased, but there would be significant dependence between multiple frequency projections.
This will be addressed and dealt with in the frequency synthesis step.

The method/model for frequency projection, of course, is not limited to simple linear regressions.
Distributed lags (e.g. beta or Almon lags) or dimension reduction methods, such as PCA \citep{bernanke2005measuring}, variable selection \citep{rovckova2016spike}, or even simple averages (in a similar fashion to bridge equations), can be extremely potent, especially when the difference in frequency is large (daily or hourly data to quarterly data, for example).
However, for relatively mild difference in frequency, e.g., monthly to quarterly, regression based projection, including dynamic linear models \citep[DLM:][]{WestHarrison1997book2,Prado2010}
\begin{align}\label{eq:proj3}
	y_t&=\bbeta_t'\z_t+\epsilon_t, \quad \epsilon_t\sim N(0,\nu_t)\\
	\bbeta_t &=\bbeta_{t-1}+\u_t, \quad \u_t\sim N(0,\nu_t\U_t) \nonumber
\end{align}
should be suitable and adequate.
The DLM specification for frequency projection is particularly appealing to applications in macroeconomics, due to its ability to capture the dynamic shifts in the underlying economy, represented in the high frequency data.
The flexibility in modeling, including thresholding for large $p$ \citep{Nakajima2010,Nakajima2011a,Nakajima2014} or stochastic volatility, can also be easily introduced, depending on the decision maker's preferences.

In the frequency projection step, the decision maker generates predictive distributions, denoted as $h_j(x_{t,j})$, for each $j=1:J$ high frequency data series.
Due to the generality of the framework, the decision maker can also opt to produce multiple $h_j(x_{t,j})$ for each $j$ with different projection specifications.
These projections form the building blocks for the succeeding synthesis step, and thus, building better, interpretable frequency projections will lead to potential improvements overall.
However, defining what constitutes a better, interpretable frequency projections is specific to the decisions and problems considered.

\subsection{Frequency synthesis}

The second step in MFS, the frequency synthesis step, then synthesizes the projections generated in the frequency projections step.
There are several points that need to be considered before proceeding with the proposed synthesis approach.
First, as stated previously, the projections, $h_j(x_{t,j})$, are biased unless the projection model correctly specifies the data generating process.
If one of the projections is the data generating process, synthesizing the projections using Bayesian model averaging (BMA) will be optimal, as the BMA weights will degenerate to the true projection with the number of observations.
However, it is seldom the case that the decision maker correctly specifies the data generating process, if that is even possible, and BMA typically degenerates to the wrong model, fast.
The projections, $h_j(x_{t,j})$, therefore, must be assumed to be biased.

Second, due to the complexity of the underlying economy, we can fully expect that these higher frequency data to be dependent.
For example, if we consider modeling quarterly GDP with monthly indicators, such as industrial production and unemployment, the assumption that the monthly indicators, as it relates to GDP, are independent to each other  is unrealistic.
 We understand and expect that the underlying economy is intertwined, particularly with variables such as industrial production and unemployment.
In the frequency projection step, each frequency is projected separately, independently; ignoring the potential dependencies among series.
The rationale to model the frequency projection separately is to allow flexibility on the decision maker's part, since ignoring the potential dependencies does give the decision maker a lot of modeling freedom.
If the decision maker wishes to use different levels of frequency or believes that certain specifications are better suited for specific data (e.g. DLM for monthly data, distributed lags for daily data), generating projections separately grants the decision maker more options, even though the dependencies are lost.
 That being said, the decision maker can always specify more complex frequency projection models that incorporate multiple high frequency data, if warranted.

The frequency synthesis step must then address these two problems: bias and dependency.
Pooling methods, including BMA, typically weigh models-- in this case high frequency data-- based on their predictive performance; mean squared error, marginal likelihood, etc.
While these methods are useful for nowcasting exercises given some specific utility, it does not give the decision maker insight into the biases and latent dependencies within and between high frequency data.
This poses serious problems with any method that attempts to combine/synthesize without taking into account these biases and dependencies, as this will lead to biased, and often incorrect, weights that might severely skew inference, forecasts, and decisions.

To effectively exploit and utilize data sampled at different frequencies-- learning their biases and latent dependencies-- to improve forecasts/nowcasts, we turn to the BPS framework of \cite{McAlinn2016}.
The BPS framework of \cite{McAlinn2016} lays out the foundational developments in expert opinion analysis \citep{GenestSchervish1985,West1992c,West1992d} and applies it to a broader framework of combining multiple predictive distributions.
In the BPS framework, a Bayesian decision maker aims to predict an outcome $y$ using information from $J$ individual projections (in this case, projections from different frequencies), where the projections are the nowcast information in terms of their pdfs, $h_j(x_j).$   
The collection of $h_j(x_j)$ defines the information set $\mH = \{ h_1(\cdot), \ldots, h_J(\cdot) \}$, which is then {\it synthesized} using the implied posterior $p(y|\mH)$. 

Employing this framework, the frequency synthesis step synthesizes nowcasts coming from the frequency projection step, where the projections are treated as latent factors of the high frequency data linked to the low frequency target data.
Therefore, the biases and latent dependencies of the high frequency projections are learnt via its posterior, which makes this framework particularly suitable for mixed frequency analysis.
The next section lays out the pertinent dynamic extension of MFS and exact specification used in the application of nowcasting U.S. GDP growth.

\subsection{Dynamic MFS}

Mixed frequency analyses are intrinsically dynamic.
While using dynamic models have proven successful in macroeconomics \citep{Cogley2005,Primiceri2005,Koop2009,Nakajima2010}, many mixed frequency models, in particular MIDAS and MF-VAR, have suffered limitations.
The MFS specifications presented in this paper exploits the underlying dynamics of the economy, realized through data from different frequencies, by utilizing DLMs for frequency projection and synthesis.

Let us consider a situation where the decision maker is sequentially forecasting, or nowcasting, a time series $y_t, t=1,2, \ldots.$
At each time point in the available frequency, she receives projections from each high frequency series.
The projections are generated using DLMs of the form in eqn.~(\ref{eq:proj3}), so the projections reflect the dynamic relationship between frequencies, this is crucial sine we expect gradual and/or dramatic changes to take place over time.
At each time $t-1,$ the decision maker aims to nowcast $y_t$ and receives current projection densities $\mH_t = \{ h_{t1}(x_{t,1}),\ldots, h_{tJ}(x_{tJ}) \}$ from the set of high frequency data. 
The full information set used, at time $t-1$ forecasting time $t$, is thus  $\{ y_{\seq1{t-1}}, \ \mH_{\seq1t} \}.$ 
 As time passes, and as the decision maker receives more information, she learns about relationships among series and their nowcast and dependency characteristics.
The  resulting Bayesian model will thus reflect this process and involve parameters that define the MFS framework and for which she updates information over time. 
Following the dynamic extension of BPS by \cite{McAlinn2016}, the decision maker has a time $t-1$ distribution for $y_t$ of the form, 
\begin{equation}\label{eq:theorem}
	p(y_t|\bPhi_t,y_{\seq1{t-1}},\mH_{\seq1t}) 
	\equiv p(y_t|\bPhi_t,\mH_t)=\int \alpha_t(y_t|\x_t,\bPhi_t)
		\prod_{j=\seq1J}h_{tj}(x_{tj})dx_{tj}
\end{equation}
where $\x_t=x_{t,\seq1J}$ is a $J-$dimensional latent projection state vector at time $t$, 
$\alpha_t(y_t|\x_t,\bPhi_t)$ is the decision maker's conditional calibration  pdf for $y_t$ given $\x_t,$ 
and $\bPhi_t$ represents time-varying parameters defining the calibration pdf-- parameters 
for which $\mD$ has current beliefs represented in terms of a current
(time $t-1$) posterior $p(\bPhi_t|\y_{\seq1{t-1}},\mH_{\seq1{t-1}}).$  
The above formulation provides a theoretically coherent framework for Bayesian prior-posterior updating, where the projections are treated as prior distributions.
This allows great flexibility on the part of the decision maker to decide what best reflects her view on how the projections to be synthesized and how they may be biased and dependent.

If the decision maker wants to reflect her belief that different frequency data interact with the quantity of interest dynamically (i.e. changing over time), a dynamic regression for the synthesis function, $\alpha_t(y_t|\x_t,\bPhi_t)$, is likely suitable.
Under this setup, we might have the following
\begin{equation} \label{eq:BPSnormalalphadynamic}
\alpha_t(y_t|\x_t,\bPhi_t) = N(y_t|\F_t'\btheta_t,v_t) \quad\textrm{with}\quad \F_t=(1, \x_t')' \quad\textrm{and}\quad
\btheta_t=(\theta_{t0},\theta_{t1},...,\theta_{tJ})',
\end{equation} 
as the synthesis function.
The latter being the $(1+J)-$vector of time-varying bias/calibration coefficients and the conditional variance $v_t$ defining residual variation beyond that explained by the regression on latent projections.   
Explicitly, the functional model parameters are now $\bPhi_t = (\btheta_t,v_t)$ at each time $t.$ 
This MFS specification defines the first component of the standard conjugate form dynamic linear model~\citep[Section 4]{WestHarrison1997book2}
\begin{subequations}
\label{DLM}
\begin{align}
	y_t&=\F_t'\btheta_t+\nu_t, \quad \nu_t\sim N(0,v_t), \label{eq:DLMa} \\
	\btheta_t&=\btheta_{t-1}+\bomega_t, \quad \bomega_t\sim N(0, v_t\W_t)\label{eq:DLMb}
\end{align}
\end{subequations}
where  $\btheta_t$ evolves  in time according to a linear/normal random walk with innovations variance matrix $v_t\W_t$ at time $t$,  and $v_t$ is the residual variance in predicting $y_t$ based on past information and the set of projection nowcast distributions.
The residuals $\nu_t$ and evolution {\em innovations} $\bomega_s$ are independent over time and mutually independent for all $t,s.$
Of course, more complex dynamic synthesis functions can be considered, depending on the decision making process and utility.

For the MFS application considered in this paper, we further specify features to the model that will further make the synthesis function more suitable.
We first allow for simple form of stochastics to make the model more flexible and sensitive to time-varying change.
This is done by using standard discount factor based methods, long used in the core Bayesian forecasting literature~\cite[e.g.][]{WestHarrison1997book2,Prado2010} and of increasing use in econometric and financial applications in more recent times~\citep[e.g.][]{dangl2012,koop2013,GruberWest2016BA,GruberWest2017ECOSTA,ZhaoXieWest2016ASMBI}. 
The time-varying intercept and projection coefficients $\btheta_t$ follow the random walk evolution of \eqno{DLMb} where $\W_t$ is defined via a standard, single discount factor specification~(\citealt[][Sect 6.3]{WestHarrison1997book2}; \citealt[][Sect 4.3]{Prado2010}), using a state evolution discount factor $\beta\in (0,1].$  
The residual variance $v_t$ follows a standard beta-gamma random walk volatility model~(\citealt[][Sect 10.8]{WestHarrison1997book2}; \citealt[][Sect 4.3]{Prado2010}), with $v_t = v_{t-1}\delta/\gamma_t$ for some discount factor $\delta\in (0,1]$ and where $\gamma_t$ are beta distributed innovations, independent over time and independent of $\nu_s,\bomega_r$ for all $t,s,r$.
Given conjugate priors on the parameters and the discount factors, we have a full synthesis function for MFS.

Under the BPS framework, the $\x_t$ vectors in each $\F_t$ are considered to be latent variables defining a dynamic latent factor model in eqns.~(\ref{DLM}).
Consider at time $t-1,$ the set of projection densities becomes available for nowcasting $y_t;$  then, from the BPS foundation,  each $x_{tj}$ is a latent draw from $h_{tj}(\cdot).$
With eqns.~(\ref{eq:DLMa},\ref{eq:DLMb}),  we have 
\begin{equation}\label{eq:dfmh}
 p(\x_t| \bPhi_t,\y_{\seq1{t-1}},\mH_{\seq1t}) \equiv p(\x_t|\mH_t) = \prod_{j=\seq1J} h_{tj}(x_{tj}) 
\end{equation}
for all time $t$ and with $\x_t,\x_s$ conditionally independent for all $t\ne s.$ 
Note that the independence of the $x_{tj}$ {\em conditional on the $h_{tj}(\cdot)$} does not equate to the decision maker's modeling and estimation of the {\em dependencies among projections};  this is critically reflected through the  effective DLM parameters $\bPhi_t = (\btheta_t,v_t).$

This is a critical point of using the BPS framework for MFS as this formulation  learns the biases and inter-dependencies over time.
As an example, consider a conditionally normal example, where $(\x|y)$ is normal with mean $\bmu+\bbeta y$ and covariance matrix $\V$.
Here $\bmu$ and $\bbeta$ account for agent-specific biases and $\V$ the inter-dependencies between agents.
Of course, to the decision maker, $\bmu$, $\bbeta$, and $\V$ are unknown and only observes $p(\x|\a,\A)$, where $\A$ is diagonal.
Through the posterior updating of BPS, $(\x|y)$ is updated, learning the biases and inter-dependencies, which is then theoretically mapped onto the effective calibration parameters, $\bPhi_t$, in the synthesis function.
Therefore, the implied prior on these effective parameters is all that is needed to synthesize the agent densities, even though the the information may come indirectly through priors on $(\bmu,\bbeta,\V)$.

It is also important to note that the above dynamic synthesis function is not the only possible way to synthesize mixed frequency projections.
For example, \cite{McAlinn2016} shows how many other model combination techniques, such as BMA, are a special case of BPS.
As such, different synthesis functions should be used if the decision maker believes it is warranted.
If the decision maker believes one of the higher frequency to be the true data generating process, a BMA style approach would be better.
Alternatively, if the decision maker believes higher frequency data interact with the low frequency data through nonlinear interactions, a nonlinear synthesis function, such as a log-linear combination, might be useful.
This flexibility in the synthesis specification grant the decision maker to better reflect her belief of how these data interact.
Interestingly, MIDAS regressions can also be considered as a special case as well.
Consider the simple MIDAS regression in \cite{ghysels2004midas}:
\begin{equation}\label{eq:midas}
y_t = \beta_0+\beta_1B(L^{1/n})z_{t-1}^{(n)}+\epsilon_{t}^{(n)},\quad \epsilon_{t}^{(n)}\sim N(0,\nu)
\end{equation}
where $B(L^{1/n})$ is a polynomial in the $L^{1/n}$ operator and $L^{m/n}z_t=z_{t-m/n}$.
As such, we can set $B(L^{1/n})z_{t-1}^{(n)}$ as $h(x_t)$, where $h(x_t)$ is a polynomial function of a series of $z_{t-m/n}$, and $\alpha(y_t|x_t)=N(y_t|\beta_0+\beta_1x_t,\nu)$.
Given $h(x_t)$ is a point estimate, the integral in eqn.~\ref{eq:theorem} collapses and we have eqn.~\ref{eq:midas} as a result.
We note that, while $h(x_t)$ is not exactly a projection of high frequency data to low frequency data in this case, the BPS framework does allow for this flexibility.

\subsection{Nowcasting with Leads \label{sec:lead}}

The biggest appeal of utilizing mixed frequency data is in its leading information.
In macroeconomic contexts, certain indicators sampled between the low frequency data  may contain critical, up-to-date information about the indicator of interest.
For example, large shocks in the financial market may happen between a quarter.
One of the purposes of mixed frequency analysis is thus to utilize this information to update and revise between data samples.

A typical approach to utilize leading information is direct projection, as with MF-VAR. 
In MF-VAR, the VAR parameters are updated and the missing low frequency data interpolated as the model receives new data (for example, every month).
The direct approach follows traditional updating and nowcasting via simulation as for 1-step ahead without leads. 
In other words, the model used for leads and no leads is the same, and everything is continuously updated (either in a DLM setting or extending/rolling window).
The straightforward approach for MFS is to similarly update the frequency synthesis parameters, as with traditional DLMs.
While this is theoretically correct, it fails to update/calibrate based on the leading information, relying wholly on the model as fitted, which can be problematic if certain data have different information relevancy over time.

\cite{McAlinn2016} has shown that customized multi-step forecasting to outperform direct projection.
This involves a  modification  in which the model at time $t-m/n$ for predicting $y_t$ changes as follows.  
With leading information, modify the MFS so that the projections made at time $t-m/n$  replace $h_{t,j}(x_{t,j})$ in  the resulting model analysis. 
This changes the interpretation of the dynamic model parameters $\{ \btheta_t, v_t\}$ to be explicitly geared to the amount of leading information. 
Bayesian model fitting then naturally \lq\lq tunes" the model to the available leading information. 

 While this typically does provide superior predictive performance, due to its calibration towards the different horizons, one cost, of course, is that a bank of MFS models is now required for any set of horizons of interest; that is, different models will be built for each release of high frequency series, so increasing the computational effort required. 
 This can be problematic, since this requires the decision maker to create and fit new models for potentially every release timing.
However, the MFS computation is relatively fast and running multiple models simultaneously should not be much of a burden on modern computers.
 
\subsection{MFS Computation \label{sec:comp}}

The process of MFS can be broken up into two components: frequency projection and frequency synthesis.
The computation related to frequency projection is rather straightforward.
For $j=1{:}J$ number of high frequency time series, we simply fit each series to the low frequency quantity of interest, separately.
If we specify a linear projection, this can be done simply by OLS, or Bayesian linear regression.
Naturally, a more elaborate projection, such as polynomials can be used.
Here, however, we believe a DLM in eqn.~\ref{eq:proj3} will be most adequate, which is easily estimated using the Kalman filter (\citealt{WestHarrison1997book2}; \citealt{Prado2010}).

The posterior computation in the frequency synthesis step requires more care.
In general the MCMC procedure requires two main blocks: estimating  the parameters in the synthesis function and estimating the latent states from the high frequency projections.
By recursively sampling from the two blocks, we effectively learn the biases and inter-dependencies of the frequency projections and then update the synthesis parameters accordingly.
We first initialize by sampling $\x_t$ from the priors $h_\ast(\ast).$
As the synthesis function used here is also a DLM, the first block simply samples from the conditional posterior of the DLM dynamic parameters $\bPhi_{\seq1t}$, given samples from the latent projections $\x_{\seq1t}$ and data  $\y_{\seq1t}$.
Thus, samples of the new parameters from $p( \bPhi_{\seq1t} |  \x_{\seq1t}, \y_{\seq1t} )$ using the standard forward filtering, backward sampling (FFBS) algorithm (e.g.~\citealt{Schnatter1994}; \citealt[][Sect 15.2]{WestHarrison1997book2}; \citealt[][Sect 4.5]{Prado2010}).

Then, given the new samples from $p( \bPhi_{\seq1t} |  \x_{\seq1t}, \y_{\seq1t} )$, we can draw from the new projection states from $ p( \x_{\seq1t} |  \bPhi_{\seq1t}, \y_{\seq1t}, \mH_{\seq1t} ).$
It is immediate that the $\x_t$ are conditionally independent over time $t$ in this conditional distribution, with time $t$ conditionals $$ p( \x_t|  \bPhi_t, y_t, \mH_t) \propto N(y_t|\F_t'\btheta_t, v_t) \prod_{j=\seq1J} h_{tj}(x_{tj}) 
	\quad\textrm{where}\quad  \F_t=(1, x_{t1},x_{t2},...,x_{tJ})'.$$
	In our application, we consider  projection nowcast densities that are T,  which can be represented as a scale mixture of normals, yielding a scale mixture multivariate normal for $\x_t$ that is trivially sampled. 
The above formulation is illustrative of how the posterior samples come from a joint distribution of latent projection states (with possible dependence among states), even if we receive the projections independently, as indicated by the independence in the right hand side.
In cases where projections are not scale mixtures of normal, either a Metropolis-Hastings simulator or an augmentation method can be employed to sample from the conditional posterior.

Repeating the two-block MCMC procedure for frequency synthesis until convergence, provides the the synthesis parameters $\bPhi_{\seq1t}$.
For nowcasting/forecasting, we utilize each sample from $\bPhi_{t}$ to draw $v_{t+1}$ from its discount volatility evolution model, and then  $\btheta_{t+1}$ conditional on $\btheta_t,v_{t+1}$ from the evolution model~\eqno{DLMb}.
This provides a draw  $\bPhi_{t+1} = \{ \btheta_{t+1}, v_{t+1} \}$ from $p(\bPhi_{t+1} |\y_{\seq1t}, \mH_{\seq1t} )$.
We can then draw $\x_{t+1}$ independently from the updated frequency projection $h_{t+1,j}(x_{t+1,j}),$  $(j=\seq1J)$.
Combining the two, we can draw $y_{t+1}$ from the conditional normal of~ \eqno{DLMa} given these sampled parameters and projection states. 
With enough samples, we generate the nowcast distribution for time $t+1$.
Detailed computation for the dynamic MFS used for the application in Section~\ref{nowGDP} is available in Appendix~\ref{supp:comp}.

\section{Nowcasting GDP\label{nowGDP}}

\begin{figure}[htbp!]
\centering
\includegraphics[width=1\textwidth]{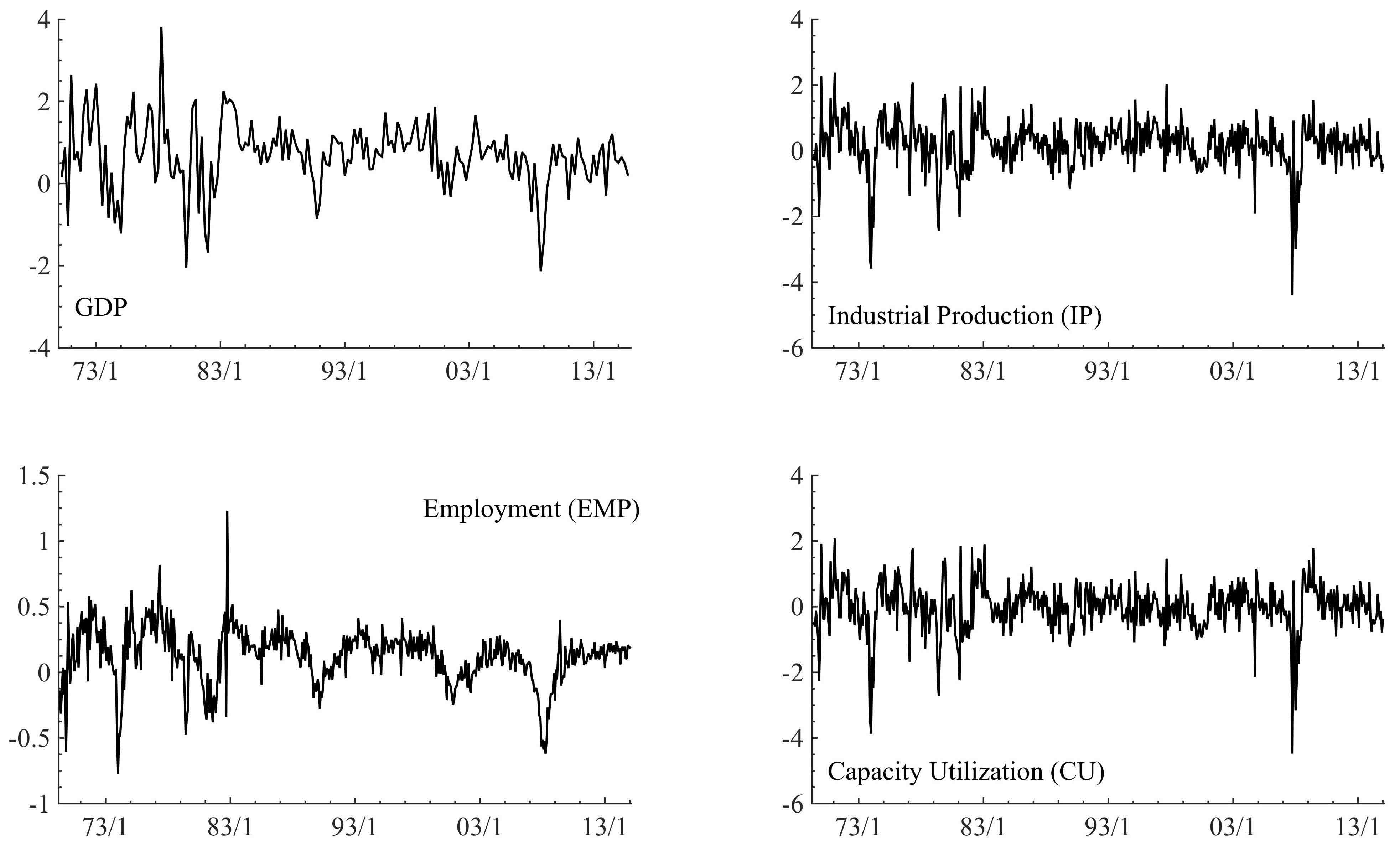} 
\caption{US macroeconomic series 1991/Q1-2015/Q4: Quarterly and monthly US macroeconomic time series (indices $\times$100 for $\%$ basis).
\label{mfs:data}}
\end{figure}

To exemplify and highlight our proposed method, we analyze quarterly U.S. macroeconomic data, focusing on nowcasting U.S. GDP with
1-quarter ahead interests using monthly U.S. macroeconomic data.
The study involves one quarterly macro series and three monthly macro series (Fig.~\ref{mfs:data}):  GDP (quarterly), industrial production (IP; monthly), employment (EMP; monthly), and capacity utilization (CU; monthly) in the U.S. economy from 1970/1 to 2015/12, a context that has received particular interest~\citep{ClementsGalvao2014,Aastveit2014,Aastveit2015}.
We focus on nowcasting GDP using  past values of itself and the higher frequency predictors underlying a set of four time series models-- the $J=4$ projections--  to be evaluated, calibrated, and synthesized.   
The time frame analyzed here is of special interest due to a number of shocks and regime changes that happened.
For example,  the early 1990s recession, the Asian and Russian financial crises in the late 1990s, the dot-com bubble in the early 2000s, and the sub-prime mortgage crisis and great recession of the late 2000s.
These shocks to the US economy are evident during this period, testing the predictive ability of any models and strategies under pressure. 
The purpose of mixed frequency analysis in this context is to utilize inter-quarter information in order to update and revise nowcasts of GDP.
As such, an effective and useful mixed frequency strategy must be able to capitalize on the leading information during shocks and regime changes.

In our analysis, we conduct a horse race between MFS and three popular methods used in the literature.
The first is a simple AR$(3)$ DLM on the quarterly GDP, using no high frequency data.
The second is the unrestricted MIDAS regression with AR$(0,1,3)$  \citep{ForoniMarcellinoSchumacher2013}, for each of the three high frequency series (we also used Beta and Almon lags, but found the unrestricted lags to be overall superior).
We then combine the individual MIDAS models using equal weight averaging.
The third is the MF-VAR model of \cite{Aastveit2014}, where a mixed frequency analysis with bootstrap is employed for each of the three high frequency series.
As with the MIDAS regression, these models for each high frequency series is then combined using equal weight averaging.
We note that while MIDAS regressions were fit with a ten year rolling window, MF-VAR was estimated in an extending window, due to estimation instability with smaller window sizes.

For the analysis of GDP nowcasting using MFS, we have four projections:
AR$(3)$, a time-varying auto-regressive model with lag 3 for GDP at the quarterly period $t$;
CU$(3)$, a dynamic model with 3-month lagged vales of capacity utilization; 
EMP$(3)$, a dynamic model with 3-month lagged vales of employment; 
IP$(3)$, a dynamic model with 3-month lagged vales of industry production.
If we do not have any leads, the latter three models will include data from [$t$, $t-2/3$, $t-1/3$].
When monthly data are available, such as two months into a quarter, the latter three models will include data from [$t+2/3$, $t+1/3$, $t$].

Each model is fit using a DLM in eqn.~(\ref{eq:proj3}).
Prior specifications for the DLM state vector and discount volatility model in each is based on $\btheta_0|v_0\sim N(\zero, v_0\I)$ and $1/v_0\sim G(1, 0.01)$, using the $(\btheta,v)$ DLM notation~\citep[][Chap 4]{WestHarrison1997book2}. 
Each projection model uses standard discount factor $(\beta)$ specification for state evolution variances and discount factor $(\delta)$ for residual volatility; we use $(\beta,\delta)=(0.99,0.95)$ in each of these projection models.

In the dynamic MFS model, for leads and without leads, we take initial priors as $\btheta_0\sim N(\m, \I)$ with $\m=(0,\one'/p)'$,  and $1/v_0\sim G(5, 0.01)$. 
Both discount factors for MFS is based on $(\beta,\delta)=(0.95,0.97)$. 
The discount factor specification is based on its performance during the testing period.
While generally discount factors should be near one, using the testing period to decide the discount factors is advisable and recommended.
Furthermore, discount factors can be estimated, though this will add a computation step within the MCMC procedure.

The analysis is done by splitting the data into three periods.
In the first period, the projection models are analyzed in parallel over 1970/Q1-1978/Q2 as a training period.
This period is used to calibrate the frequency projections, in order to obtain good estimates.
In the second period, from 1978/Q3-1990/Q4, the frequency projections are continuously being updated and calibrated, and we begin our estimation of MIDAS, MF-VAR, and MFS.
The last period, from 1991/Q1-2015/Q4, we continuously and sequentially update and nowcast quarterly GDP, for both no leads and with leads of two months (i.e. utilizing data two months into the quarter).
During this period, we compare the predictive performances of the models considered.
This testing period spans over a quarter century, and we are able to explore predictive performance over periods of drastically varying economic circumstances, check robustness, and compare benefits and characteristics of each strategy.
Predictive performances are compared using two metrics; point and density predictive metrics.
For point nowcasts, we compute and compare mean squared nowcast error (MSNE).
A traditional metric used in these studies that reflect a simple squared error loss utility.
For density nowcasts, we evaluate using the log predictive density ratios (LPDR).
This is computed for each leading period and across time indices $t$, as 
\begin{align*}
	\mathrm{LPDR}_{\seq1t}(1)=\sum_{i=\seq1t}\mathrm{log}\{p_{s}(y_{t+1}|y_{1{:}t})/p_{\mathrm{MFS}}(y_{t+1}|y_{1{:}t})\},
\end{align*}
where $p_s(y_{t+1}|y_{1{:}t})$ is the predictive density under the models competing with MFS.
LPDR measures, as used by several authors recently~\cite[e.g.][]{Nakajima2010,Aastveit2015,McAlinn2016}, provide a direct statistical assessment of relative accuracy at multiple horizons that extend traditional 1-step focused Bayes' factors.
The LPDR goes beyond MSFE and weighs and compares dispersion of nowcast densities along with location, to elaborate on raw MSNE measures.
This is quite critical with the increased importance and need for density nowcasts/forecasts in the field of macroeconomics.


\section{Results and Comparison \label{resGDP}}

Comparing predictive summaries over the out-of-sample period, MFS improves 
nowcasting accuracy relative to the AR$(3)$ model, all MIDAS regressions, and all MF-VARs by at least 10$\%$; see numerical summaries in Table~\ref{table:mfs}.
Focusing on the point nowcast, excluding MFS, we see that AR$(3)$ outperforms MIDAS for no leads, except for the equal wight average of MIDAS without AR factors as well as all MF-VARs.
While this is somewhat puzzling, as we expect models with mixed frequency data to do better than a simple model with only the low frequency data, this might be explained by the dynamics captured by the DLM for AR$(3)$, a feature not incorporated in both MIDAS and MF-VAR.
We also note that MF-VARs mostly underperforms compared to MIDAS, which is consistent with the findings of \cite{kuzin2011midas,foroni2014comparison}.
 
With incorporating leading information, we note that most of the MIDAS regressions and MF-VARs outperform AR$(3)$, except for using capacity utilization.
This is consistent with previous research, as we expect MIDAS and MF-VAR to perform better with leading information into the quarter, updating and revising the status of the economy with new, up-to-date information.
We also find that, consistently, capacity utilization to underperform, suggesting that employment and industry production to be better providers of inter-quarter information for nowcasting GDP.
Additionally, contrary to \cite{kuzin2011midas,foroni2014comparison}, MF-VARs do tend to outperform MIDAS overall with leading information.

In contrast to MIDAS and MF-VAR, MFS is able to achieve superior point nowcasts by  synthesizing the information from the higher frequency data.
MFS significantly outperforms all of the other models, with increases of 14-70$\%$, when we include leading information.
Additionally, MFS with leads outperforms itself with no leads by roughly 15$\%$, demonstrating how synthesizing inter-quarter high frequency data can improve nowcasting.
This is an important and significant point, as the purpose of mixed frequency analysis is to improve predictive performance by capitalizing on the information provided through inter-quarter data.

These results are echoed for the distribution nowcasts as well, as MFS is available to ascertain better measures of uncertainty from the high frequency data, with consistent improvements over all MIDAS regressions and MF-VARs.
A couple of points are particularly noteworthy.
For one, for the most part, MIDAS and MF-VARs seem to do worse in terms of LPDR compared to the simple AR$(3)$ model, which is consistent with the point nowcast performances.
Apart from MF-VAR, averaging MIDAS regressions seem to decrease distributional nowcast accuracy.
Another point is that MFS improves with leading information over itself as well, as indicated by the difference between the LPDR of AR$(3)$.
This further heightens the potency of MFS when incorporating leading information.

\begin{figure}[htbp]
\centering
\includegraphics[width=0.75\textwidth]{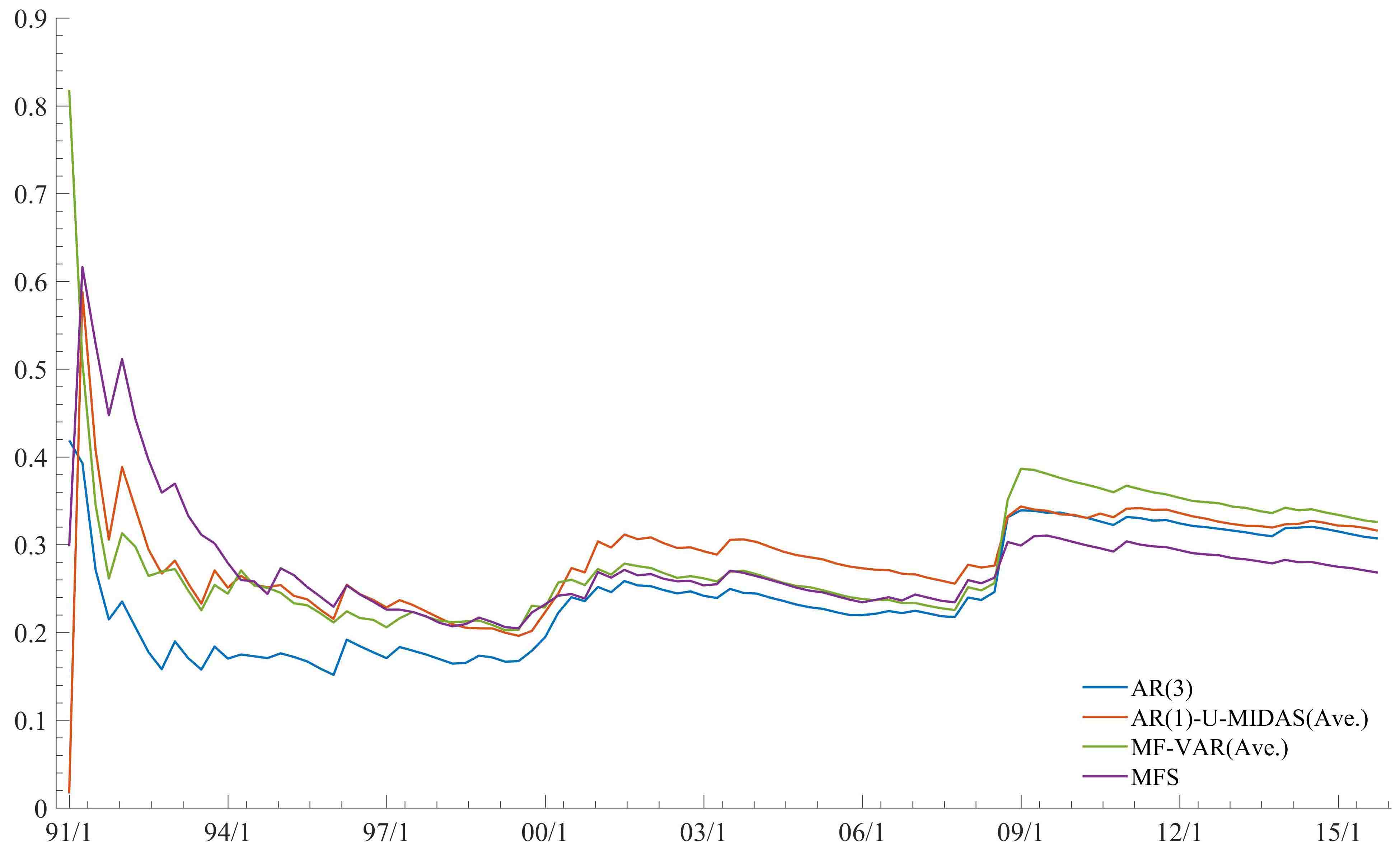} 
\caption{US GDP nowcasting 1991/Q1-2015/Q4: Mean squared 1-quarter ahead nowcast errors MSNE$_{1{:}t}(1)$ of quarterly GDP, sequentially revised at each of the $t=\seq1{100}$ quarters with no leads. 
\label{mfs:3mse}}
\end{figure}

\begin{figure}[htbp]
\centering
\includegraphics[width=0.75\textwidth]{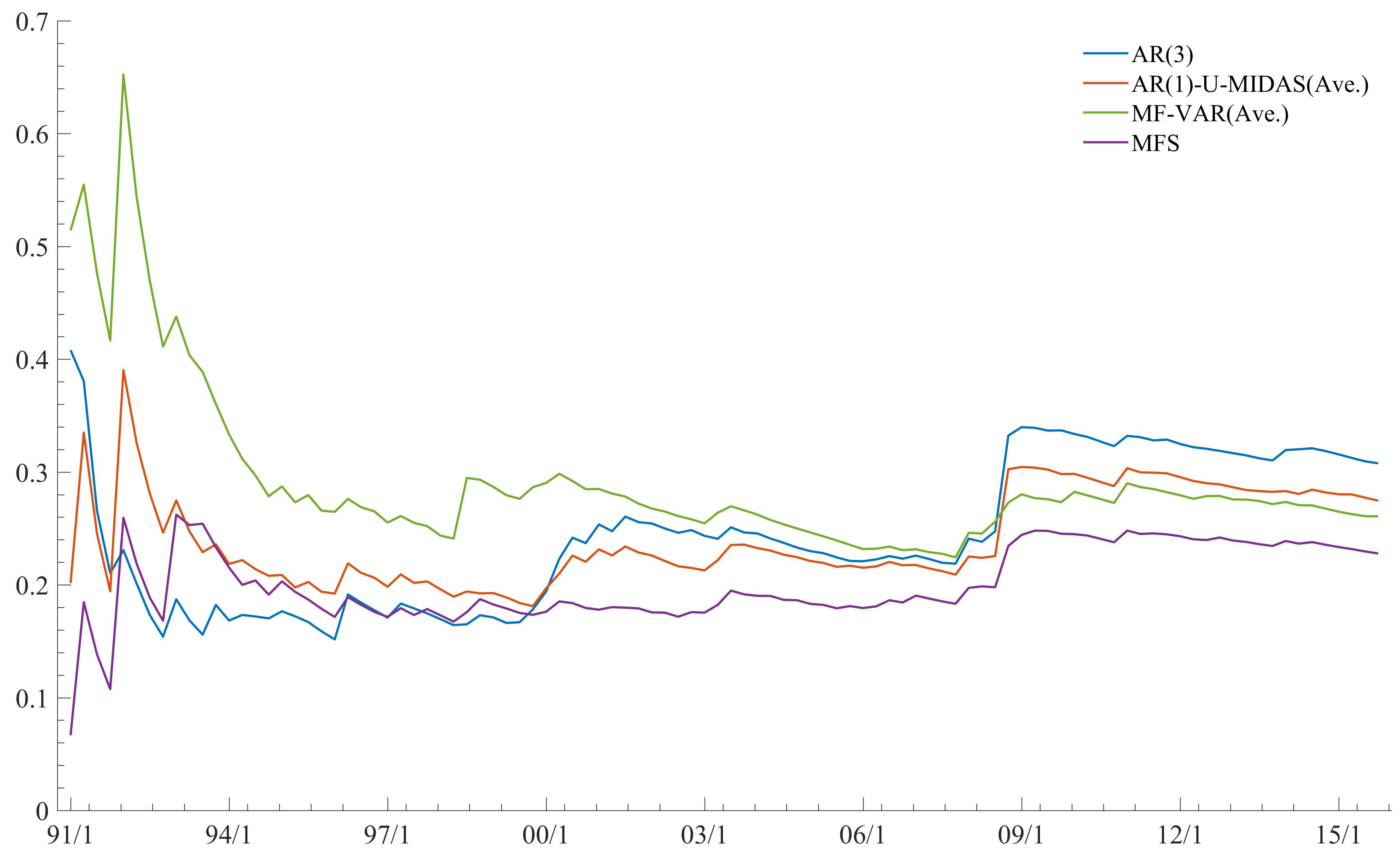} 
\caption{US GDP nowcasting 1991/Q1-2015/Q4: Mean squared 1-month ahead nowcast errors MSNE$_{1{:}t}(1)$ of quarterly GDP, sequentially revised at each of the $t=\seq1{100}$ quarters with leads of two months.  
\label{mfs:1mse}}
\end{figure}

We further our analysis by reviewing summary graphs showing aspects of analyses evolving over time 
during the testing period, a period that includes challenging economic times that impede good predictive performance. 
Figs.~\ref{mfs:3mse} and~\ref{mfs:1mse} show the 1-step ahead measures of MSNE$_{1{:}t}(1)$ for each time $t$ for no leads and two month leads.
Note that we only compare against equal weight averaged AR$(1)$-U-MIDAS, averaged MF-VAR, and AR$(3)$, for the sake of legibility.
For no leads, we see that AR$(3)$ outperforms the rest up until the sub-prime mortgage crisis, but MFS eventually surpasses AR$(3)$.
While MFS, has some problems at the beginning of the analysis, which is somewhat expected since MSNEs are cumulative and earlier periods unstable, MFS quickly improves over time with consistent gains in predictive performance.
In terms of MIDAS and MF-VAR, there is a clear shift in performance before and after the crisis.
While MF-VAR does seem to outperform MIDAS before the crisis, it appears that the crisis causes significant loss in predictive power in MF-VAR, compared to MIDAS.

With leads, MFS outperforms the rest throughout most of the period of analysis.
Note that the shock during the sub-prime mortgage crisis is mitigated in MFS, while the AR$(3)$ and MIDAS suffer greatly as they poorly adapt.
On the other hand, while MF-VAR does suffer a great amount at the beginning and most of the first half of the analysis,  it appears to be effected by the crisis the least and comparable to MFS.
Overall, MFS improves over all models considered, but it is interesting that MF-VAR seems to be incorporating high frequency data the best out of the models considered during the crisis.

\begin{figure}[b]
\centering
\includegraphics[width=0.75\textwidth]{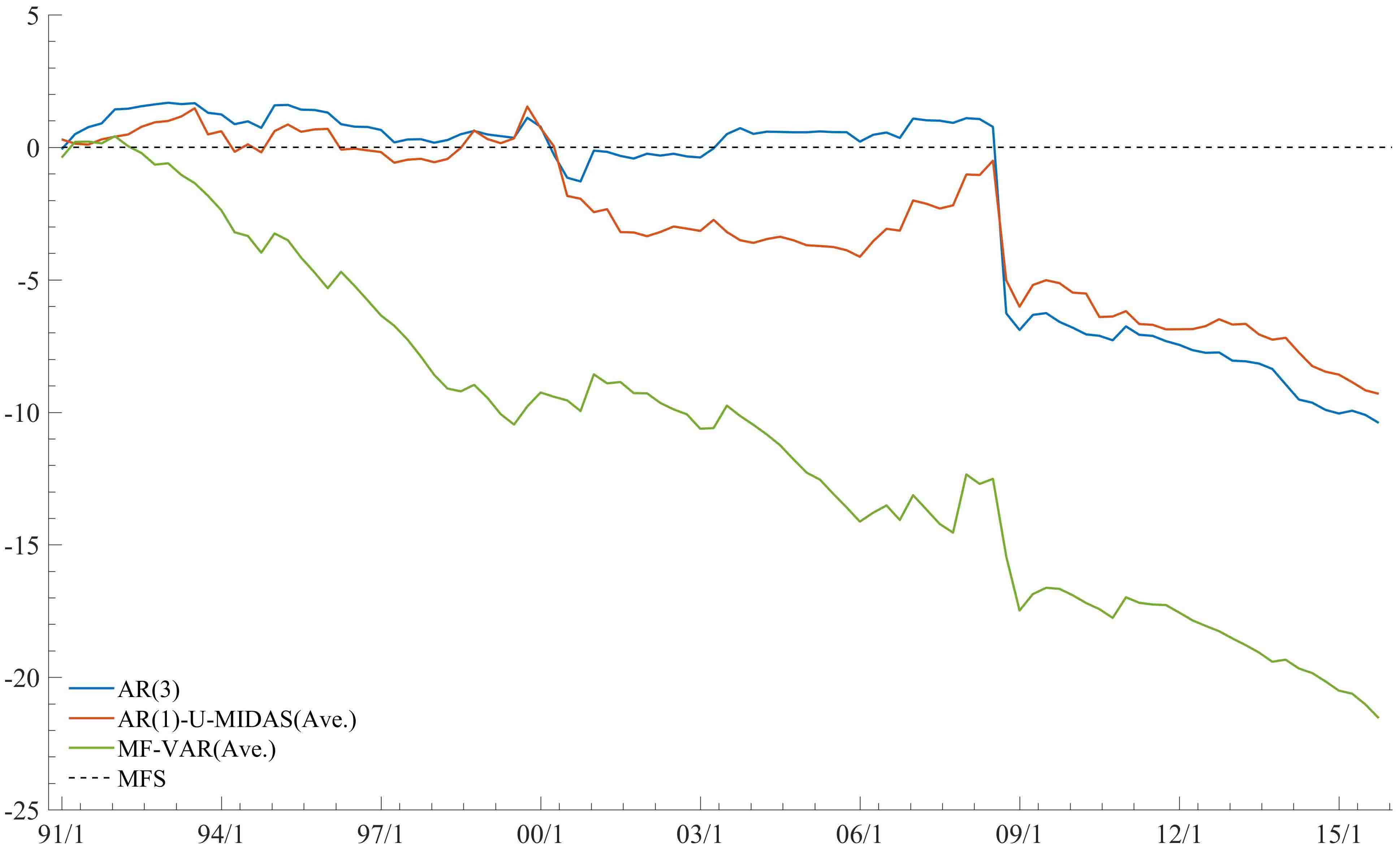} 
\caption[US GDP nowcasting 1991/Q1-2015/Q4: 1-quarter ahead log predictive density ratios LPDR$_{1{:}t}(1)$ of quarterly GDP, sequentially revised at each of the $t=\seq1{100}$ quarters with no leads.]{US GDP nowcasting 1991/Q1-2015/Q4: 1-quarter ahead log predictive density ratios LPDR$_{1{:}t}(1)$ of quarterly GDP, sequentially revised at each of the $t=\seq1{100}$ quarters with no leads. The baseline at 0 over all $t$ corresponds to the standard MFS model. 
\label{mfs:3lpdr}}
\end{figure}
\begin{figure}[htbp]
\centering
\includegraphics[width=0.75\textwidth]{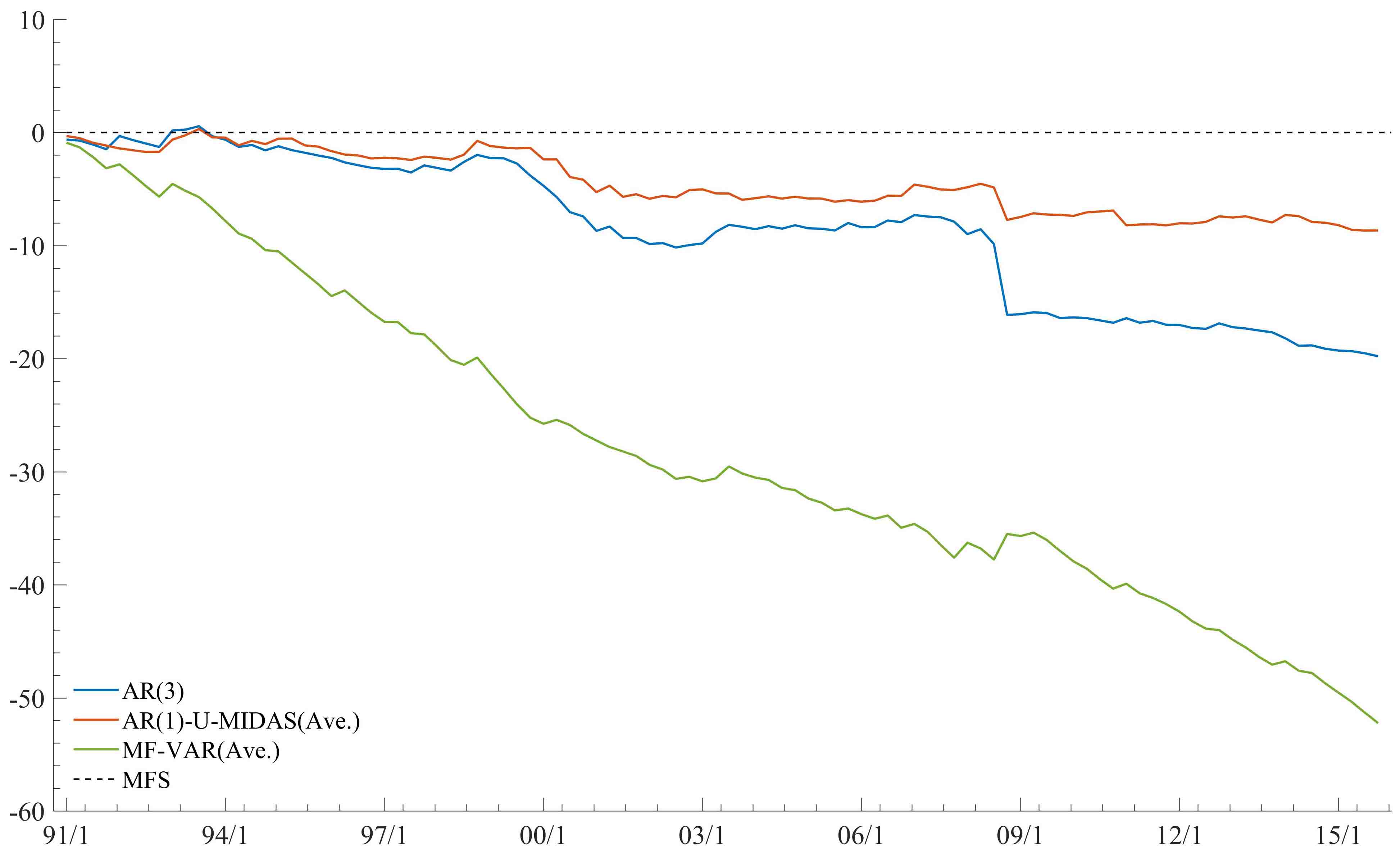} 
\caption[US GDP nowcasting 1991/Q1-2015/Q4: 1-month ahead log predictive density ratios LPDR$_{1{:}t}(1)$ of quarterly GDP,  sequentially revised at each of the $t=\seq1{100}$ quarters with leads of two months.]{US GDP nowcasting 1991/Q1-2015/Q4: 1-month ahead log predictive density ratios LPDR$_{1{:}t}(1)$ of quarterly GDP, sequentially revised at each of the $t=\seq1{100}$ quarters with no leads. The baseline at 0 over all $t$ corresponds to the standard MFS model. 
\label{mfs:1lpdr}}
\end{figure}

The distribution nowcast results, Figs.~\ref{mfs:3lpdr} and \ref{mfs:1lpdr}, echo the point nowcast results.
When there are no leads, we see MFS, AR$(3)$, and MIDAS to perform fairly similarly until the crisis.
However, after the crisis, there is a large drop in LPDR, followed by a gradual decrease afterwards.
MF-VAR, on the other hand, sees a continuous decrease over time.
During crisis periods, we see significant decrease in LPDR for AR$(3)$ and MIDAS (and to a lesser extent MF-VAR).
Amplifying the results with no leads, results with leads strengthens the position of MFS; with MFS almost always outperforming the models compared.
As LPDRs are relative, these results demonstrate how the uncertainty assessment of MFS (additional to the mean estimates) is continuously and accurately measured and robust under shocks, with high frequency data providing sequential information representing the current economic situation.
This is prevalent-- and most critical-- for MFS with leads, as shocks can happen within a quarter that cannot be picked up by an AR$(3)$ model only using quarterly data.

\begin{figure}[htbp]
\centering
\includegraphics[width=0.75\textwidth]{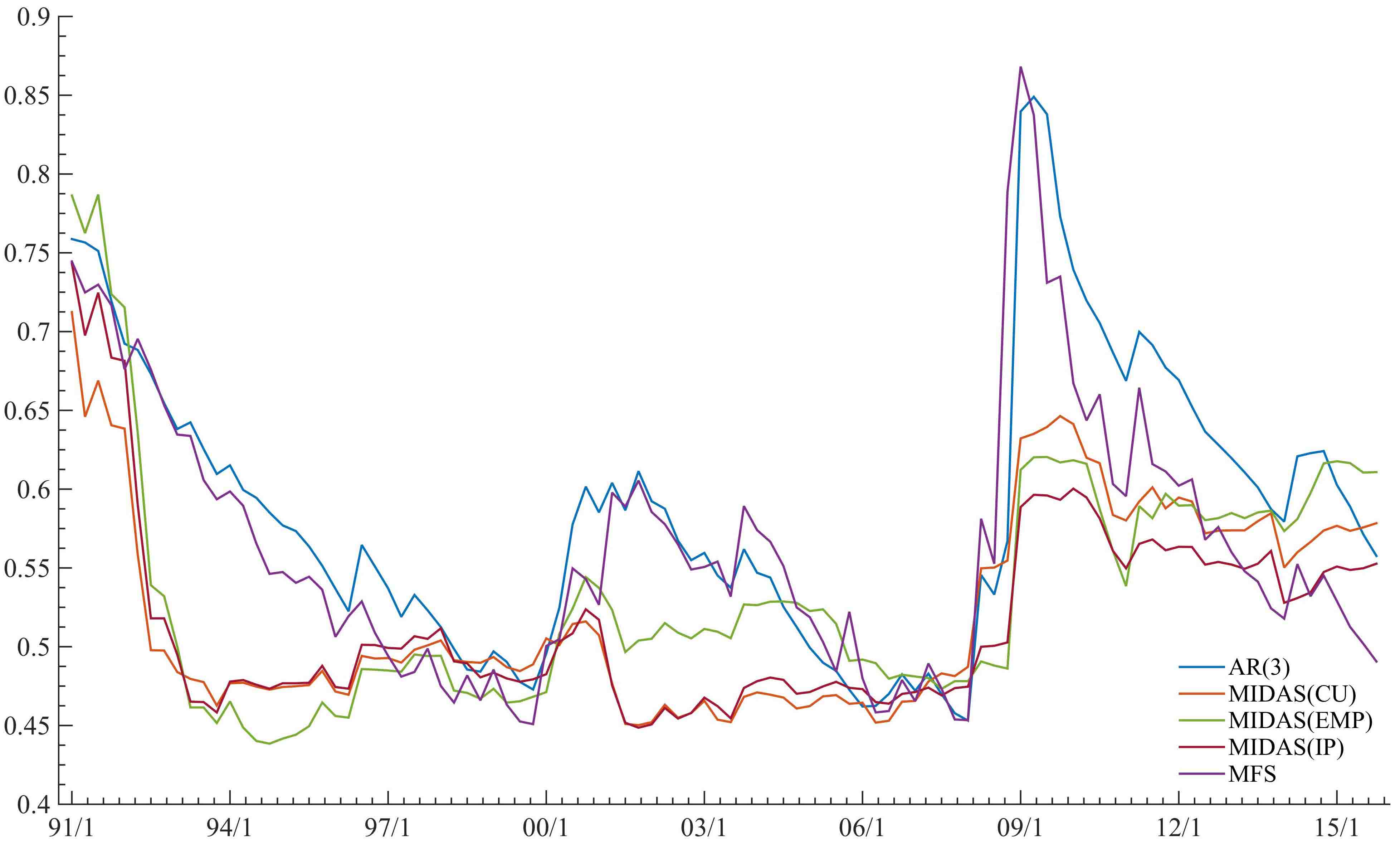} 
\caption{US GDP nowcasting 1991/Q1-2015/Q4: 1-step ahead nowcast standard deviations
sequentially computed at each of the $t=\seq1{100}$ quarters with no leads. 
\label{mfs:3sd}}
\end{figure}

\begin{figure}[htbp]
\centering
\includegraphics[width=0.75\textwidth]{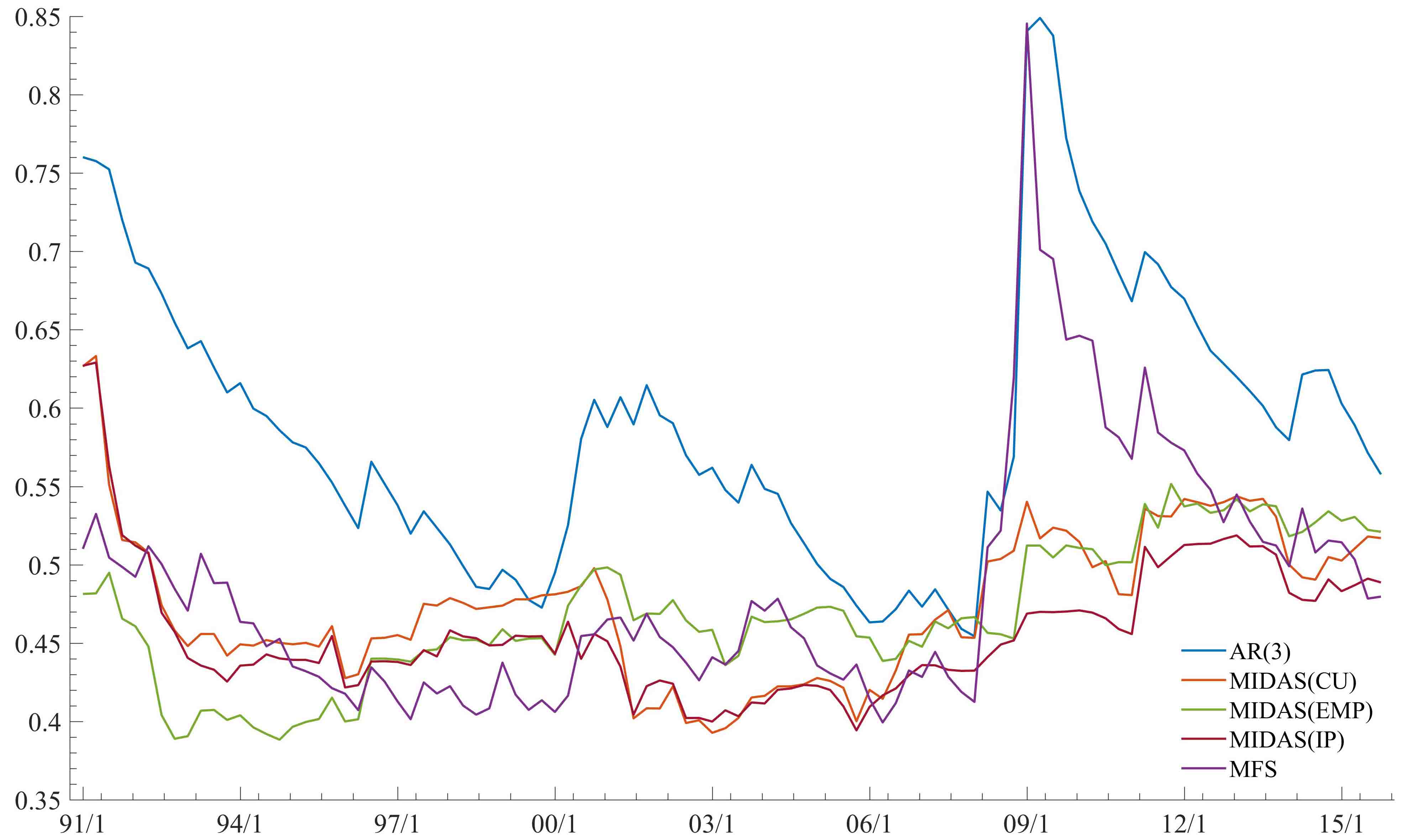} 
\caption{US GDP nowcasting 1991/Q1-2015/Q4: 1-step ahead nowcast standard deviations
sequentially computed at each of the $t=\seq1{100}$ quarters with leads of two months. 
\label{mfs:1sd}}
\end{figure}

Figs.~\ref{mfs:3sd} and~\ref{mfs:1sd} compare the 1-step ahead standard deviation results.
Here, we compare MFS with AR$(3)$ and different AR$(3)$-U-MIDAS specifications.
Without leads, MFS nowcast standard deviations seem to mirror that of the AR$(3)$, though with some notable differences during and after the crisis.
Comparatively, MIDAS models have smaller standard deviations throughout, even during the crisis; a setback due to the restriction of MIDAS having static volatility.
MFS and MIDAS both have lower standard deviations when we include leads, reflecting refined filtered nowcasts throughout a quarter, where the standard deviation profile of MFS being much more similar to that of MIDAS.
However, the limitations of MIDAS having static volatility hinders adaptability during the crisis.
MFS, on the other hand, is able to adapt dynamically, as seen in the jump in standard deviation around 2009.
This flexibility in volatility is reflected through the LPDR results.

\begin{figure}[t]
\centering
\includegraphics[width=0.75\textwidth]{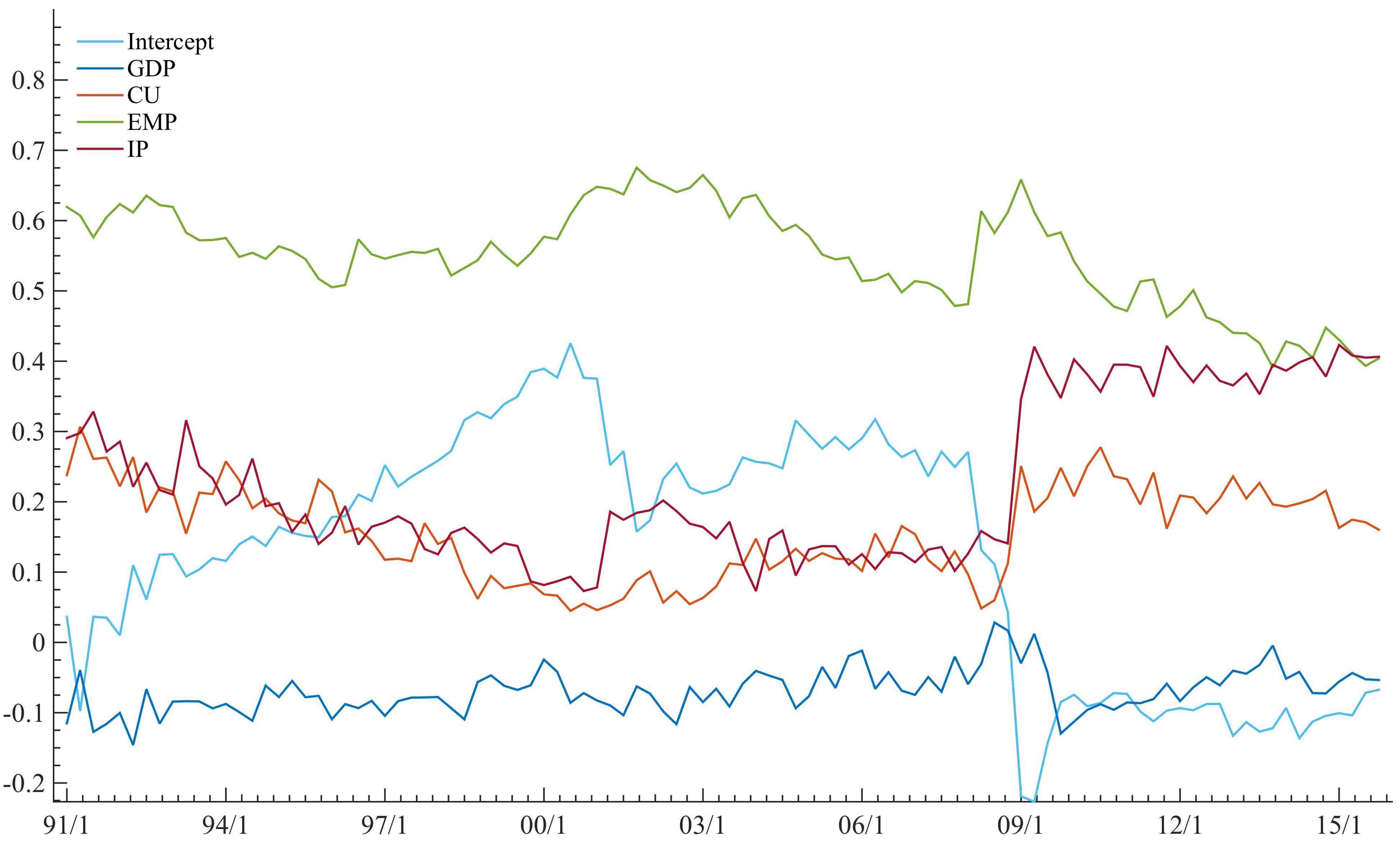} 
\caption{US GDP nowcasting 1991/Q1-2015/Q4:  On-line posterior means of MFS model
 coefficients sequentially computed at each of the $t=\seq1{100}$ quarters with no leads. 
\label{mfs:3coef}}
\end{figure}
\begin{figure}[htbp]
\centering
\includegraphics[width=0.75\textwidth]{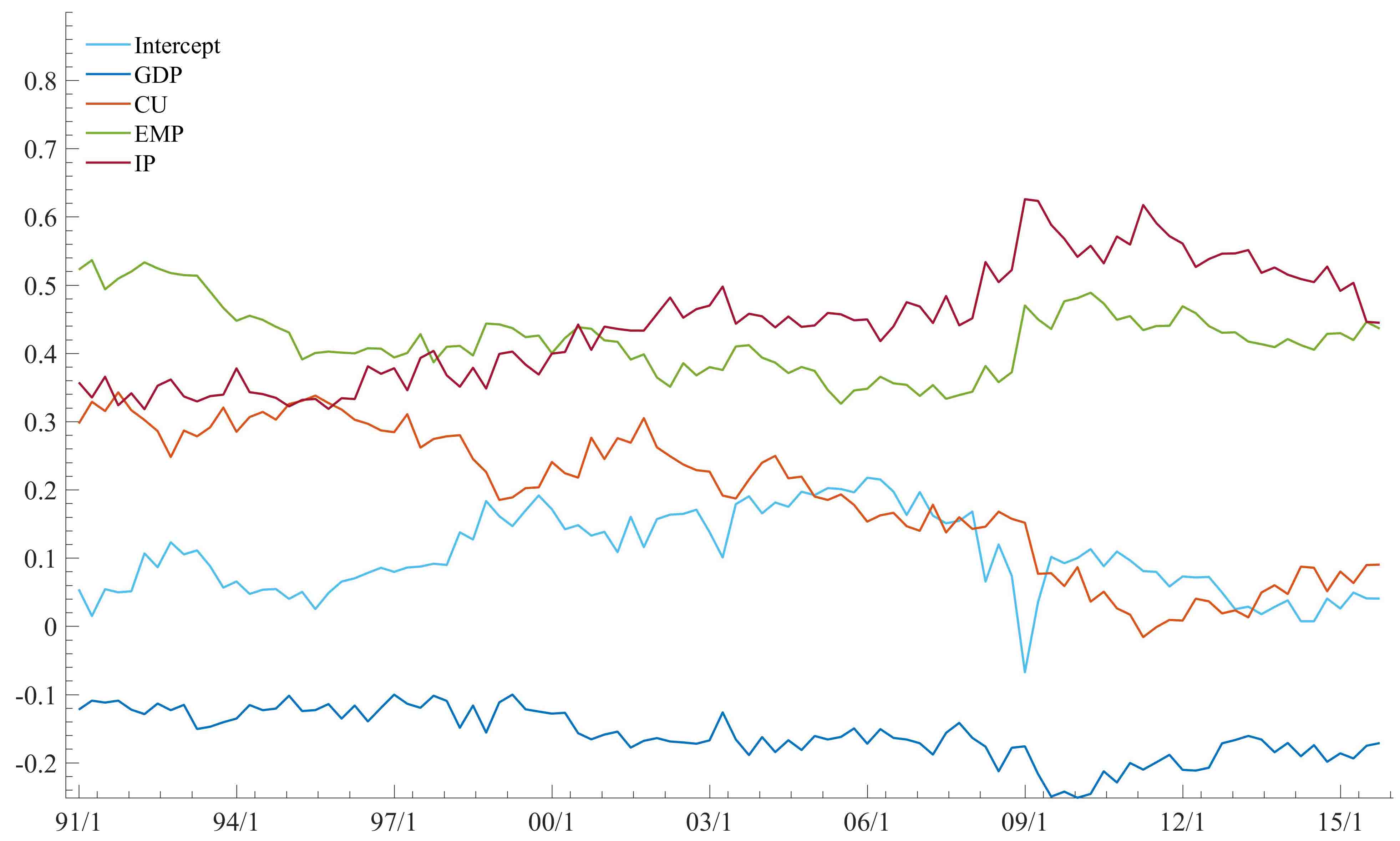} 
\caption{US GDP nowcasting 1991/Q1-2015/Q4:  On-line posterior means of MFS model
 coefficients sequentially computed at each of the $t=\seq1{100}$ quarters with leads of two months. 
\label{mfs:1coef}}
\end{figure}

Moving on to MFS on-line posterior means of MFS coefficients (Figs.~\ref{mfs:3coef} and~\ref{mfs:1coef}), we get a better understanding on how MFS improves nowcasts by dynamically adapting the high frequency information.
For the nowcasts with no leads (Fig.~\ref{mfs:3coef}),  the information coming from employment is the most significant.
During the crisis, there is a significant increase in information provided by industrial production, perhaps because industrial production was quicker to adapt to the shock compared to employment and capacity utilization.
Interestingly, GDP (AR$(3)$) provides very little information in terms of model coefficients.
This can be seen as a lack of persistence in the lags of GDP, with more information coming from economic indicators that effect GDP rather than GDP of past.
This is particularly useful for policy makers making decisions based on metrics they can control, such as interest rates, which are often sampled at higher frequency.

For nowcasts with leads (Fig.~\ref{mfs:1coef}), there is less variability during the shock, with more of a persistent structural change over time.
Like that of nowcasts with no leads, employment starts out providing the most information, then is later taken over by industrial production at around the dot-com bubble.
Capacity utilization decreases over time in an almost inverse proportion to industrial production.
This can be seen as the economy shifting from a more manufacturing economy to a more IT based economy, with the persistent decline of the industrial sector having a larger effect on the economy as a whole.

\begin{figure}[htbp]
\centering
\includegraphics[width=0.75\textwidth]{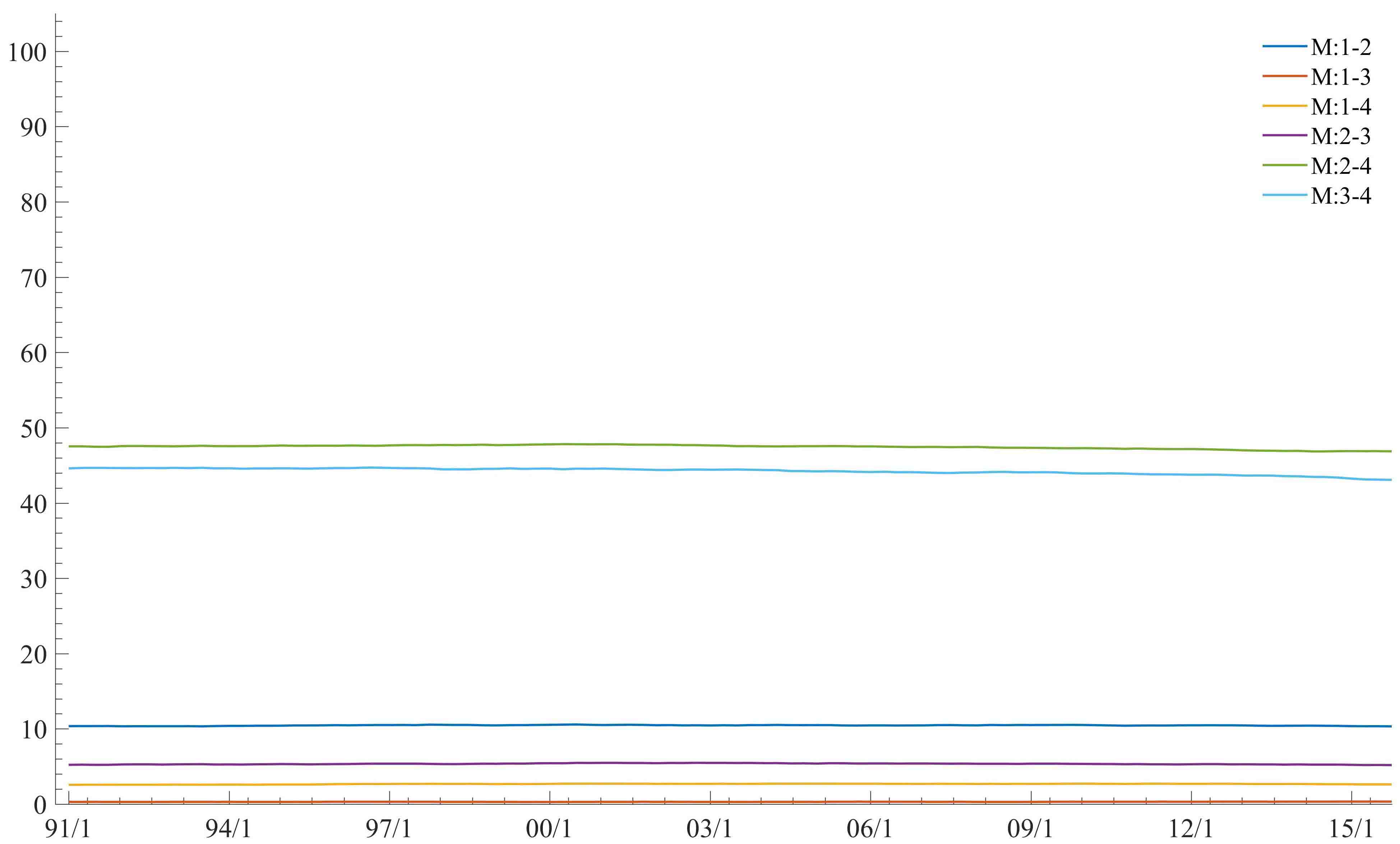} 
  \caption{US GDP nowcasting 1991/Q1-2015/Q4: MFS (no leads) model-based trajectories  of paired MC-empirical R$^2$  in the posterior for the latent agent states $x_{jt}$  for $j=\seq14$ over the $t=\seq1{100}$ quarters.}
    \label{3r2per}
\end{figure}

\begin{figure}[htbp]
\centering
\includegraphics[width=0.75\textwidth]{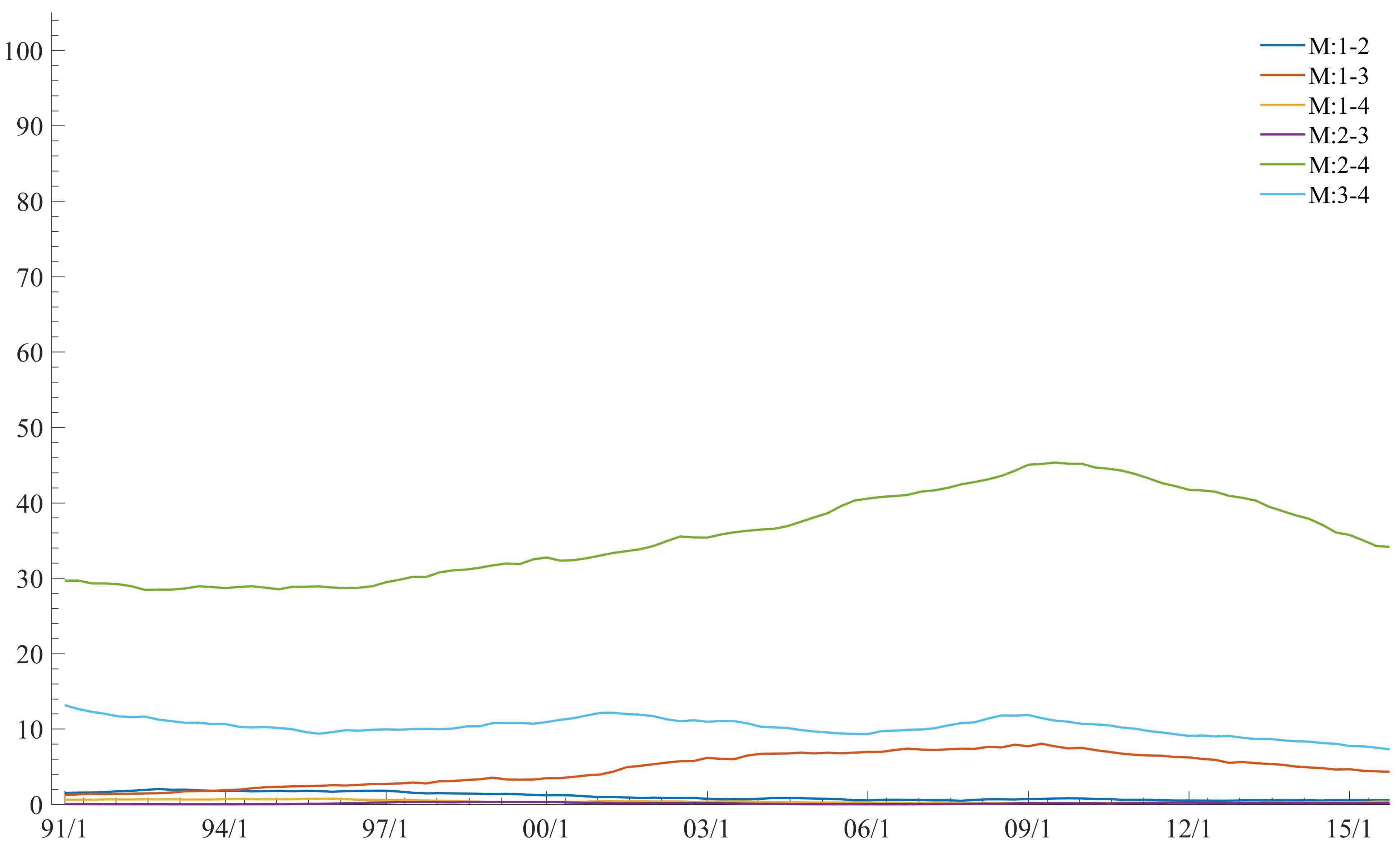} 
  \caption{US GDP nowcasting 1991/Q1-2015/Q4: MFS (two month lead) model-based trajectories  of paired MC-empirical R$^2$  in the posterior for the latent agent states $x_{jt}$  for $j=\seq14$ over the $t=\seq1{100}$ quarters.}
    \label{1r2per}
\end{figure}

Finally, we investigate the posterior for $\x_{\seq1T}$ in order to infer on the latent dependencies  between projections.  This is not of standard form and
is represented in terms of the MCMC-based posterior sample.   A simple, though insightful, summary is to look at the empirical R$^2$ measures, which is computed by the MCMC approximate posterior 
variance matrix of $\x_t$ at each $t,$ and from that extract implied sets of conditionals variances of
all pairs of $x_{tj}$ given another $x_{ti}, i\ne j.$    We do this  using each single agent $i\ne j$, defining paired  MC-empirical R$^2$ measures of how dependent agents $i,j$ are-- the {\em bivariate dependencies}.
Fig.~\ref{3r2per} and~\ref{1r2per} displays trajectories over time for these two measures from the  MFS analysis with and without leads. 
 
Overall, we see relatively low paired conditional dependencies for with and without leads.
Notably, without leads, the conditional dependencies are stable, with almost no movement over time, and the dependence with IP with both CU and EMP being higher than the rest.
With leads, we observe significant dynamics in the dependence structure.
As with no leads, IP and CU have the highest dependency followed by IP and EMP, though the latter is significantly lower than with leads.
The dependence between IP and CU, most interestingly,  gradually increase, peaked at around the subprime mortgage crisis, and then dropping off.
Similarly, GDP and EMP displays a gradual increase peaked at around the subprime mortgage crisis, though to a lesser degree to IP and CU.
We finally note that the dependency with leads is overall lower than without leads.
This perhaps reflects the filtering nature of information; as data accrues within a quarter, each information coming from the projects are made more distinct (filtered), thus dependencies and uncertainties about them made clearer, resulting in lower paired dependencies.



\section{Summary \label{sum}}
Drawing on theory and framework of Bayesian predictive synthesis, 
we develop a theoretically and conceptually sound framework to analyze mixed frequency data.
With this new framework, policy makers are able to dynamically  learn and update nowcasts using data sampled at different frequencies, especially with leading information. 

The nowcasting study of U.S. GDP illustrates how effective and practical MFS is under settings that are increasingly important and topical in macroeconomics and econometrics.
Using the two stage process of MFS, it improves nowcast performance and dominates other standard strategies, for nowcasting with and without leads, and for both point and distribution forecasts.
Further analysis shows evidence that, using leading information, MFS is extremely robust in its forecast abilities under economic distress, which is critically important for practical applications.
Additionally, posterior inference of the full time series provides the policy maker with  information on how agents are related, 
and how that relationship dynamically evolves over time.

There are multiple extensions that can be considered for Bayesian MFS.
In particular, extensions into  data that are sampled at significantly higher frequency is of interest.
This includes daily, or tick, data sampled in finance, where, per quarter, we might observe hundreds or thousands of data points.
This demands development into different specifications in the frequency projection step in order to deal with the ``$p>>t$" problem.
Using distributed lags, shrinkage priors, or thresholding should be effective under these settings.
Furthermore, the MFS framework allows for development of multivariate synthesis functions, which allows for analyzing multiple low frequency data using multiple high frequency data, simultaneously, potentially improving nowcasts and inference.

%
%

\bibliographystyle{elsarticle-harv}
\bibliography{References}

\newpage

\begin{center}
{\Large Dynamic Mixed Frequency Synthesis\\ for Economic Nowcasting} 

\bigskip
{\large Kenichiro McAlinn} 

\bigskip
{\Large  Tables and Figures} 

\bigskip\bigskip
\end{center}

\begin{table}[htbp!]
\caption[US GDP nowcasting 1991/Q1-2015/Q4: Nowcast evaluations for quarterly US GDP.]{US GDP nowcasting 1991/Q1-2015/Q4: Nowcast evaluations for quarterly US GDP, 
 comparing mean squared nowcast errors and log predictive density ratios for this $T=100$ quarters. Note: LPDR$_{1:T}$ is relative to MFS and $t+\#/3$ denotes the number of monthly leads into the next quarter.}
\centering
\begin{tabular}{lrrrrrr}
                                  & \multicolumn{3}{c}{$t$}                                                        & \multicolumn{3}{c}{$t+2/3$}                                          \\
\multicolumn{1}{l|}{}             & MSNE$_{1:T}$        & \%                   & \multicolumn{1}{r|}{LPDR$_{1:T}$} & MSNE$_{1:T}$        & \%                   & LPDR$_{1:T}$        \\ \hline
\multicolumn{1}{l|}{AR(3)}        & 0.3071               & $-$14.50               & \multicolumn{1}{r|}{$-$10.40}        & 0.3071               & $-$34.79               & $-$19.79               \\
                                  & \multicolumn{1}{l}{} & \multicolumn{1}{l}{} & \multicolumn{1}{l}{}               & \multicolumn{1}{l}{} & \multicolumn{1}{l}{} & \multicolumn{1}{l}{} \\
     \multicolumn{1}{l}{MIDAS}                                                               & \multicolumn{1}{l}{} & \multicolumn{1}{l}{} & \multicolumn{1}{l}{}               & \multicolumn{1}{l}{} & \multicolumn{1}{l}{} & \multicolumn{1}{l}{} \\
\multicolumn{1}{l|}{Unrestricted} &                      &                      &       \multicolumn{1}{l|}{}                             &                      &                      &                      \\ \hline
\multicolumn{1}{l|}{CU}           & 0.3369               & $-$25.62               & \multicolumn{1}{r|}{$-$11.55}        & 0.3537               & $-$55.74               & $-$27.65               \\
\multicolumn{1}{l|}{EMP}          & 0.3479               & $-$29.71               & \multicolumn{1}{r|}{$-$14.93}        & 0.2837               & $-$24.52               & $-$13.81               \\
\multicolumn{1}{l|}{IP}           & 0.3126               & $-$16.54               & \multicolumn{1}{r|}{$-$8.97}        & 0.2959               & $-$29.87               & $-$19.15               \\
\multicolumn{1}{l|}{Ave.}         & 0.3019               & $-$12.56               & \multicolumn{1}{r|}{$-$6.89}        & 0.2666               & $-$17.03               & $-$5.69               \\
                                  & \multicolumn{1}{l}{} & \multicolumn{1}{l}{} & \multicolumn{1}{l}{}               & \multicolumn{1}{l}{} & \multicolumn{1}{l}{} & \multicolumn{1}{l}{} \\
\multicolumn{1}{l|}{Unr.+AR(1)}   &                      &                      &         \multicolumn{1}{l|}{}                           &                      &                      &                      \\ \hline
\multicolumn{1}{l|}{CU}           & 0.3324               & $-$23.93               & \multicolumn{1}{r|}{$-$11.38}        & 0.3696               & $-$62.25               & $-$30.35               \\
\multicolumn{1}{l|}{EMP}          & 0.3567               & $-$33.00               & \multicolumn{1}{r|}{$-$16.88}        & 0.2975               & $-$30.58               & $-$17.20               \\
\multicolumn{1}{l|}{IP}           & 0.3318               & $-$23.70               & \multicolumn{1}{r|}{$-$12.45}        & 0.3035               & $-$33.23               & $-$20.81               \\
\multicolumn{1}{l|}{Ave.}         & 0.3159               & $-$17.78               & \multicolumn{1}{r|}{$-$9.30}        & 0.2746               & $-$20.54               & $-$8.66               \\
                                  & \multicolumn{1}{l}{} & \multicolumn{1}{l}{} & \multicolumn{1}{l}{}               & \multicolumn{1}{l}{} & \multicolumn{1}{l}{} & \multicolumn{1}{l}{} \\
\multicolumn{1}{l|}{Unr.+AR(3)}   &                      &                      &           \multicolumn{1}{l|}{}                         &                      &                      &                      \\ \hline
\multicolumn{1}{l|}{CU}           & 0.3420               & $-$27.51               & \multicolumn{1}{r|}{$-$13.05}        & 0.3938               & $-$72.88               & $-$36.76               \\
\multicolumn{1}{l|}{EMP}          & 0.3669               & $-$36.81               & \multicolumn{1}{r|}{$-$19.05}        & 0.2977               & $-$30.70               & $-$18.90               \\
\multicolumn{1}{l|}{IP}           & 0.3368               & $-$25.58               & \multicolumn{1}{r|}{$-$12.70}        & 0.3077               & $-$35.06               & $-$22.65               \\
\multicolumn{1}{l|}{Ave.}         & 0.3280               & $-$22.29                &   \multicolumn{1}{r|}{$-$10.58}        & 0.2909               & $-$27.68               & $-$7.11               \\
                                  & \multicolumn{1}{l}{} & \multicolumn{1}{l}{} & \multicolumn{1}{l}{}               & \multicolumn{1}{l}{} & \multicolumn{1}{l}{} & \multicolumn{1}{l}{} \\
                                           \multicolumn{1}{l|}{MF-VAR}   &                      &                      &           \multicolumn{1}{l|}{}                         &                      &                      &                      \\ \hline
\multicolumn{1}{l|}{CU}           & 0.7734               & $-$188.36               & \multicolumn{1}{r|}{$-$22.85}        & 0.4517               & $-$98.29               & $-$69.50               \\
\multicolumn{1}{l|}{EMP}          & 0.3545               & $-$32.17               & \multicolumn{1}{r|}{$-$24.15}        & 0.2631               & $-$15.50               & $-$31.06               \\
\multicolumn{1}{l|}{IP}           & 0.4262               & $-$58.92               & \multicolumn{1}{r|}{$-$20.37}        & 0.2667               & $-$17.05               & $-$38.39               \\
\multicolumn{1}{l|}{Ave.}         & 0.3681               & $-$37.25                &   \multicolumn{1}{r|}{$-$21.54}        & 0.2607               & $-$14.45               & $-$50.24               \\       
                                  & \multicolumn{1}{l}{} & \multicolumn{1}{l}{} & \multicolumn{1}{l}{}               & \multicolumn{1}{l}{} & \multicolumn{1}{l}{} & \multicolumn{1}{l}{} \\                            
                                 \multicolumn{1}{l|}{MFS}   &                      &                      &                                    \multicolumn{1}{l|}{}&                      &                      &                      \\ \hline
\multicolumn{1}{l|}{}          & 0.2682                & $-$                     &  \multicolumn{1}{r|}{$-$}             & 0.2278               & $-$                    & $-$                   
\end{tabular}
\label{table:mfs}
\end{table}

\clearpage
\setcounter{page}{1}
\begin{center}
{\Large Dynamic Mixed Frequency Synthesis \\for Economic Nowcasting} 

\bigskip
{\large Kenichiro McAlinn} 

\bigskip
{\Large  Supplementary Material} 

\bigskip\bigskip
\end{center}

\appendix

\section{Appendix: Summary of MCMC for Dynamic MFS\label{supp:comp}}

\subsection{Overview and Initialization}
This appendix summarizes algorithmic details of implementation of the MCMC computations for dynamic MFS model fitted
in Section~\ref{sec:comp}.  
This involves the frequency projection step and frequency synthesis step, the former using basic, analytical DLM updating (~\citealt{WestHarrison1997book2}; \citealt{Prado2010}) and the latter using a customized two-component block Gibbs sampler used in \cite{McAlinn2016}.
Note that the MFS framework does not limit to the models explained here, and many different models can be used given the decision maker's preference and utility.

In the context of nowcasting GDP, the full MCMC analysis is performed for each quarter as we move to the next period and new data are observed.
The detailed MCMC steps here are for a specific $t$, given all data up to that point.

Standing at time $T$, the decision maker has historical information $ y_{\seq1T}$, each high frequency data  $\z_{\seq1T,j}$, for $j=\seq1J$,   initial 
priors $\bbeta_{0,j}\sim N(\m_{0,j}, \C_{0,j} \nu_{0,j}/s_{0,j} )$, $1/\nu_{0,j}\sim G(n_{0,j}/2, n_{0,j}s_{0,j}/2)$, for the frequency projection, and $\btheta_0\sim N(\m_0, \C_0 v_0/s_0 )$, $1/v_0\sim G(n_0/2, n_0s_0/2)$, for the frequency synthesis, and discount factors $(\beta_j,\delta_j)$ and $(\beta,\delta)$ for each projection models and synthesis model. 

\subsection{Frequency projection}

Given the model specification in eqn.~\ref{eq:proj3}, we generate frequency projections, $\mH_{\seq1T}$, using DLM updating.
Since the specifications in eqn.~\ref{eq:proj3} enjoys analytic conjugate updating, this is done once, for each high frequency data, separate from the two component block Gibbs sampling MCMC procedure of frequency synthesis.

\subsubsection{For each high frequency data: Updating projection distribution $h_j(x_{\seq1T,j})$ } 
Given the high frequency data  $\z_{t,j}$, we have the conjugate DLM form, 
\begin{align*}
	y_t&=\bbeta_t'\z_{t,j}+\epsilon_{t,j}, \quad \epsilon_{t,j}\sim N(0,\nu_{t,j})\\
	\bbeta_{t,j} &=\bbeta_{t-1,j}+\u_{t,j}, \quad \u_{t,j}\sim N(0,\nu_{t,j}\U_{t,j}) \nonumber
\end{align*}
with known elements $\U_{t,j}$ and specified initial prior at $t=0.$ 
Then, this follows the standard DLM updating (~\citealt{WestHarrison1997book2}; \citealt{Prado2010}).
In detail, this proceeds as follows. 

For each $t=\seq1T$ in sequence, perform the standard one-step 
filtering updates to compute and save the sequence of sufficient statistics for the on-line predictive distribution of $h_j(x_{t,j})\sim T_{n_{t,j}}(h_{t,j},H_{t,j})$ at each $t,$ for each agent $j$. 
\begin{itemize} 
	\item[FP 1.]{\em Posterior at $t-1$:} 
		\begin{align*}
		\bbeta_{t-1,j}|\nu_{t-1,j},\z_{\seq1{t-1},j},y_{\seq1{t-1}}&\sim N(\m_{t-1,j}, \C_{t-1,j}\nu_{t-1,j}/s_{t-1,j}),\\
		\nu_{t-1,j}^{-1}|\z_{\seq1{t-1},j},y_{\seq1{t-1}}&\sim G(n_{t-1,j}/2, n_{t-1,j}s_{t-1,j}/2),
		\end{align*}
		with point estimates $\m_{t-1,j}$ of $\bbeta_{t-1,j}$ and $s_{t-1,j}$ of $\nu_{t-1,j}.$ 
	\item[FP 2.]{\em Prior update to time $t$:} 
		\begin{align*}
		\bbeta_{t,j}|\nu_t,\z_{\seq1{t-1},j},y_{\seq1{t-1}}&\sim N(\m_{t-1,j}, \R_{t,j}\nu_{t,j}/s_{t-1,j})
		\quad\textrm{with}\quad \R_{t,j}=\C'_{t-1,j}/\delta_j, \\
		\nu_{t,j}^{-1}|\z_{\seq1{t-1},j},y_{\seq1{t-1}}&\sim G(\beta_j n_{t-1,j}/2, \beta n_{t-1,j}s_{t-1,j}/2),
		\end{align*}
		with (unchanged) point estimates $\m_{t-1,j}$ of $\bbeta_{t,j}$ and $s_{t-1,j}$ of $\nu_{t,j},$  but with 
		increased uncertainty relative to the time $t-1$ posteriors, the level of increased uncertainty 
			being defined by the discount factors.   
	\item[FP 3.]{\em  Compute and save projection distribution:} 
		$$y_t |\z_{\seq1t,j},y_{\seq1{t-1}} \sim T_{\beta_j n_{t-1,j}}(h_{t,j},H_{t,j})$$ where
		$$h_{t,j}=\z_{t,j}'\m_{t-1,j}\quad \textrm{and}\quad H_{t,j}=\z_{t,j}'\R_{t,j}\z_{t,j}+s_{t-1,j}.$$
	\item[FP 4.]{\em  Update to time $t$ posterior via filtering:}  
		\begin{align*}
		\bbeta_{t,j}|\nu_{t,j},\z_{\seq1{t},j},y_{\seq1{t}}&\sim N(\m_{t,j}, \C_{t,j}\nu_{t,j}/s_{t,j}),\\
		\nu_{t,j}^{-1}|\z_{\seq1{t},j},y_{\seq1{t}}&\sim G(n_{t,j}/2, n_{t,j}s_{t,j}/2),
		\end{align*}
		 with  $\m_{t,j}=\m_{t-1,j}+\A_{t,j} e_{t,j}$, $ 	\C_{t,j}=r_{t,j}(\R_{t,j}-q_{t,j}\A_{t,j}\A_{t,j}'),$ $n_{t,j}=\beta_j n_{t-1,j}+1,$ and $ s_{t,j}=r_{t,j}s_{t-1,j}$ 
		based on  the projection error  $ e_{t,j}=y_t-h_{t,j},$ the state adaptive coefficient vector 
		$\A_{t,j}=\R_{t,j}\z_{t,j}/H_{t,j},$  and volatility estimate ratio $r_{t,j}=(\beta_j n_{t-1,j}+e_{t,j}^2/H_{t,j})/n_{t,j}.$ 
\end{itemize} 

\subsection{Frequency synthesis}
Given the frequency projections, $\mH_{\seq1T}$, we then continue with the two block Gibbs sampler in \cite{McAlinn2016} for the frequency synthesis step.

As with \cite{McAlinn2016}, we first initialize by setting $\F_t=(1,x_{t1},...,x_{tJ})'$ for each $t=\seq1T$ at some chosen initial values of the
latent agent states.  
Initial values can be chosen arbitrarily. Here we simply generate agent states from their priors, i.e., 
from the agent forecast distributions,  $x_{tj} \sim h_{tj}(x_{tj})$ independently for all $t=\seq1T$ and $j=\seq1J$.
Since the projections are T, this is trivially sampled.
Note that the choice of initial values is not critical, as the MCMC is rapidly convergent.

With the initial values, the MCMC resamples from two coupled sets of conditional posteriors to generate MCMC samples from the target posterior $p(\x_{\seq1T},\bPhi_{\seq1T}|y_{\seq1T}, \mH_{\seq1T}).$
Details of the samples is as follows.

\subsubsection{Per MCMC Iterate Step 1: Sampling frequency synthesis DLM parameters $\bPhi_{\seq1T}$ } 
Conditional on any values of
the frequency projections,  we sample from a conditionally normal DLM with the frequency projections as
known predictors based on their specific values.  The frequency synthesis conjugate DLM form, 
\begin{align*}
		y_t&=\F_t'\btheta_t+\nu_t, \quad \nu_t\sim N(0,v_t), \label{eq:DLMa} \\
	\btheta_t&=\btheta_{t-1}+\bomega_t, \quad \bomega_t\sim N(0, v_t\W_t),
\end{align*}
has known elements $\F_t,\W_t$ and specified initial prior at $t=0.$   This is simulated using the efficient and standard FFBS 
algorithm, modified to incorporate the discount stochastic volatility components for $v_t$
(e.g.~\citealt{Schnatter1994}; \citealt[][Sect 15.2]{WestHarrison1997book2}; \citealt[][Sect 4.5]{Prado2010}). 
In detail, this proceeds as follows. 

\begin{itemize} 
\item[]{\em\bf Forward filtering:} For each $t=\seq1T$ in sequence, perform the standard one-step 
filtering updates to compute and save the sequence of sufficient statistics for the on-line posteriors
$p(\btheta_t,v_t|y_{\seq1t},\x_{\seq1t})$ at each $t.$ The summary technical details are as follows: 
\begin{itemize} 
	\item[FS 1.]{\em Posterior at $t-1$:} 
		\begin{align*}
		\btheta_{t-1}|v_{t-1},\x_{\seq1{t-1}},y_{\seq1{t-1}}&\sim N(\m_{t-1}, \C_{t-1}v_{t-1}/s_{t-1}),\\
		v_{t-1}^{-1}|\x_{\seq1{t-1}},y_{\seq1{t-1}}&\sim G(n_{t-1}/2, n_{t-1}s_{t-1}/2),
		\end{align*}
		with point estimates $\m_{t-1}$ of $\btheta_{t-1}$ and $s_{t-1}$ of $v_{t-1}.$ 
	\item[FS 2.]{\em Prior update to time $t$:} 
		\begin{align*}
		\btheta_{t}|v_t,\x_{\seq1{t-1}},y_{\seq1{t-1}}&\sim N(\m_{t-1}, \R_tv_t/s_{t-1})
		\quad\textrm{with}\quad \R_{t}=\C_{t-1}/\delta, \\
		v_t^{-1}|\x_{\seq1{t-1}},y_{\seq1{t-1}}&\sim G(\beta n_{t-1}/2, \beta n_{t-1}s_{t-1}/2),
		\end{align*}
		with (unchanged) point estimates $\m_{t-1}$ of $\btheta_{t}$ and $s_{t-1}$ of $v_{t},$  but with 
		increased uncertainty relative to the time $t-1$ posteriors, the level of increased uncertainty 
			being defined by the discount factors.   
	\item[FS 3.]{\em  Compute nowcast distribution:} 
		$$y_t |\x_{\seq1t},y_{\seq1{t-1}} \sim T_{\beta n_{t-1}}(f_t,q_t)$$ where
		$$f_t=\F_t'\m_{t-1}\quad \textrm{and}\quad q_t=\F_t'\R_t\F_t+s_{t-1}.$$
	\item[FS 4.]{\em  Filtering update to time $t$ posterior:}  
		\begin{align*}
		\btheta_{t}|v_{t},\x_{\seq1{t}},y_{\seq1{t}}&\sim N(\m_{t}, \C_{t}v_{t}/s_{t}),\\
		v_{t}^{-1}|\x_{\seq1{t}},y_{\seq1{t}}&\sim G(n_{t}/2, n_{t}s_{t}/2),
		\end{align*}
		 with  $\m_t=\m_{t-1}+\A_t e_t,$  $ 	\C_{t}=r_t(\R_t-q_t\A_t\A_t'),$ $n_t=\beta n_{t-1}+1,$ and $ s_t=r_ts_{t-1},$ 
		based on  nowcast error  $ e_t=y_t-f_t,$ the state adaptive coefficient vector 
		$\A_t=\R_t\F_t/q_t,$  and volatility estimate ratio $r_t=(\beta n_{t-1}+e_t^2/q_t)/n_t .$ 
\end{itemize} 
	\item[]{\em\bf Backward sampling:}  Having run the forward filtering analysis up to time $T,$ the 
	 backward sampling proceeds as follows. 
	 \begin{itemize}
	 \item[FS a.]{\em At time $T$:} 
	 
	 Simulate $\bPhi_T=(\btheta_T,v_T)$ from the final normal/inverse gamma posterior 
	  $p(\bPhi_T|\x_{\seq1{T}},y_{\seq1{T}})$ as follows. First, draw $v_T^{-1}$ from $G(n_{T}/2, n_{T}s_{T}/2),$ and then 
	  	draw $\btheta_T$ from $N(\m_T,\C_T v_T/s_T).$
	 \item[FS b.]{\em Recurse back over times $t=T-1, T-2, \ldots, 0:$}  
	 
	 At time $t,$ sample  
	 	$\bPhi_t=(\btheta_t,v_t)$ by simulating the volatility $v_t$ via 
	 			$v_t^{-1}=\beta v_{t+1}^{-1}+\gamma_t$ where $\gamma_t$ is an independent draw from
	 				$\gamma_t  \sim G((1-\beta)n_t/2,n_ts_t/2),$ then simulating the state $\btheta_t$ from the conditional normal posterior 
	 			$p(\btheta_{t}|\btheta_{t+1},v_t,\x_{\seq1T},y_{\seq1T})$ with mean 
	 			vector $\m_{t}+\delta(\btheta_{t+1}-\m_{t})$ and variance matrix 
	 			$ \C_{t} (1-\delta)(v_t/s_t).$ 	 			
	 \end{itemize} 
\end{itemize} 
\subsubsection{Per MCMC Iterate Step 2:   Sampling the latent projection states $\x_{\seq1T}$}

  Conditional on most recently
sampled values of
the frequency synthesis DLM parameters $\bPhi_{\seq1T},$   the MCMC iterate completes with resampling of the 
latent projection states from their full conditional posterior
$ p( \x_{\seq1t} |  \bPhi_{\seq1t}, y_{\seq1t}, \mH_{\seq1t} ).$    It is immediate that the $\x_t$ are
conditionally independent over time $t$ in this conditional distribution, with time $t$ 
conditionals 
\begin{equation}\label{app:ccforx}
p( \x_t|  \bPhi_t, y_t, \mH_t) \propto N(y_t|\F_t'\btheta_t, v_t) \prod_{j=\seq1J} h_{tj}(x_{tj}) 
	\quad\textrm{where}\quad  \F_t=(1, x_{t1},x_{t2},...,x_{tJ})'. \end{equation}

Since $h_{tj}(x_{tj})$ has a density of  $ T_{n_{tj}}(h_{tj},H_{tj})$,
we can express this as a scale mixture of Normal, $ N(h_{tj},H_{tj})$, with  $ \H_t=\textrm{diag}(H_{t1}/\phi_{t1},H_{t2}/\phi_{t2},...,H_{tJ}/\phi_{tJ})$, where $\phi_{tj}$ are 
independent over $t,j$ with gamma distributions, $\phi_{tj} \sim G(n_{tj}/2,n_{tj}/2).$

The posterior distribution for each $\x_t$ is then sampled, given $\phi_{tj}$, from
\begin{equation}\label{app:condx}
	p( \x_t|  \bPhi_t, y_t, \mH_t) = N(\h_{t}+\b_t c_t, \H_t-\b_t\b_t'g_t)
\end{equation}
where $c_t = y_t- \theta_{t0} - \h_t'\btheta_{t,\seq1J}$,    $g_t=v_t+  \btheta_{t,\seq1J}'\q_t\btheta_{t,\seq1J}$, and $\b_t =  \q_t\btheta_{t,\seq1J}/g_t$.
Here, given the previous values of $\phi_{tj}$, we have $\H_t=\textrm{diag}(H_{t1}/\phi_{t1},H_{t2}/\phi_{t2},...,H_{tJ}/\phi_{tJ})$
Then, conditional on these new samples of $\x_t,$  updated samples of the latent scales are drawn
from the implied set of conditional gamma posteriors $\phi_{tj}|x_{tj} \sim G((n_{tj}+1)/2,(n_{tj}+d_{tj})/2)$
where $d_{tj}= (x_{tj}-h_{tj})^2/H_{tj}$, independently for each $t,j$. 
This is easily computed and then sampled independently for each $\seq1T$ to provide resimulated agent states
over $\seq1T.$

In some cases, frequency projection densities may be more elaborate mixtures of normals or might not have analytically tractable densities and only have samples.
As discussed in \cite{McAlinn2016}, the former can be sampled via similar augmentation techniques as the T densities in this example.
For the latter, MCMC will proceed using some form of 	
  Metropolis-Hastings algorithm, or accept/reject methods, or importance sampling for the 
  latent agent states.

\end{document}